%% file: main.tex
\documentclass[fleqn,usenatbib]{mnras}
\usepackage{newtxtext,newtxmath}

\usepackage[T1]{fontenc}
\usepackage{ae,aecompl}

\usepackage{graphicx}	% Including figure files
\usepackage{amsmath}	% Advanced maths commands
\usepackage{amssymb}	% Extra maths symbols
\usepackage{pifont}     % some symbols we may want
\usepackage{xcolor}

\newcommand{\rb}{\mathrm{b}}
\newcommand{\rc}{\mathrm{c}}
\newcommand{\rd}{\mathrm{d}}
\newcommand{\re}{\mathrm{e}}
\newcommand{\ri}{\mathrm{i}}
\newcommand{\rt}{\mathrm{t}}

\newcommand{\RePart}{\mathrm{Re}}
\newcommand{\ImPart}{\mathrm{Im}}

\newcommand{\Jr}{J_{\mathrm{r}}}
\newcommand{\rra}{r_{\mathrm{a}}}
\newcommand{\rrb}{r_{\rb}}
\newcommand{\brc}{b_{\rc}}
\newcommand{\alphac}{\alpha_{\rc}}
\newcommand{\betac}{\beta_{\rc}}

\newcommand{\psiM}{\psi_\mathrm{M}}

\newcommand{\omegaM}{\omega_{\mathrm{M}}}
\newcommand{\OmegaM}{\Omega_{\mathrm{M}}}
\newcommand{\gammaM}{\gamma_{\mathrm{M}}}

\newcommand{\rapo}{r_{\mathrm{apo}}}
\newcommand{\rperi}{r_{\mathrm{per}}}
\newcommand{\alphamin}{\alpha_{\mathrm{min}}}
\newcommand{\alphamax}{\alpha_{\mathrm{max}}}

\newcommand{\br}{\mathbf{r}}
\newcommand{\brp}{\mathbf{r^{\prime}}}

\newcommand{\bn}{\mathbf{n}}
\newcommand{\bJ}{\mathbf{J}}

\newcommand{\bOmega}{\mathbf{\Omega}}

\newcommand{\vnm}{v_{\mathbf{n}}^{-}}
\newcommand{\vnp}{v_{\mathbf{n}}^{+}}
\newcommand{\omeganm}{\omega_{\mathbf{n}}^{\mathrm{min}}}
\newcommand{\omeganp}{\omega_{\mathbf{n}}^{\mathrm{max}}}

\newcommand{\calL}{\mathcal{L}}

\newcommand{\tE}{\tilde{E}}
\newcommand{\tEth}{\tilde{E}_{\mathrm{th}}}
\newcommand{\tEinit}{\tilde{E}_{\mathrm{init}}}

\newcommand{\p}{\partial}
\newcommand{\half}{\tfrac{1}{2}}

\newcommand{\boldmatrix}[1]{\boldsymbol{\mathsf{#1}}}
\newcommand{\bM}{\boldmatrix{M}}
\newcommand{\bG}{\boldmatrix{G}}
\newcommand{\bI}{\boldmatrix{I}}
\newcommand{\bvareps}{\boldmatrix{\varepsilon}}

\newcommand{\intLb}{{\int_{\substack{ \\[0.83ex] -1 \\ \hspace{-1.9em} \calL}}^{1}}}

\newcommand{\float}{\texttt{Float64}}

\newcommand{\integer}{\texttt{Int64}}
\newcommand{\bool}{\texttt{Bool}}

\newcommand{\julia}{\texttt{julia}}

\newcommand{\plummerplus}{\texttt{PlummerPlus}}
\newcommand{\EXP}{\texttt{EXP}}
\newcommand{\orbitalelements}{\href{https://github.com/michael-petersen/OrbitalElements.jl}{\texttt{OrbitalElements.jl}}}
\newcommand{\linearresponse}{\href{https://github.com/michael-petersen/LinearResponse.jl}{\texttt{LinearResponse.jl}}}
\newcommand{\fht}{\href{https://github.com/michael-petersen/FiniteHilbertTransform.jl}{\texttt{FiniteHilbertTransform.jl}}}
\newcommand{\astrobasis}{\href{https://github.com/michael-petersen/AstroBasis.jl}{\texttt{AstroBasis.jl}}}

\usepackage{acronym}

\newacro{DF}{distribution function}
\newacroplural{DF}[DFs]{distribution functions}
\newcommand{\DF}{\ac{DF}}
\newcommand{\DFs}{\acp{DF}}

\newacro{ILR}{inner Lindblad resonance}
\newcommand{\ILR}{\ac{ILR}}

\newacro{ROI}{radial orbit instability}
\newcommand{\ROI}{\ac{ROI}}

\defcitealias{Fouvry.Prunet.2022}{FP22}
\begin{document}
\label{firstpage}
\pagerange{\pageref{firstpage}--\pageref{lastpage}}

\title[Linear Response Toolbox for self-gravitating  discs and spheres]
{Predicting the linear response of self-gravitating stellar spheres\\ and discs with {\tt LinearResponse.jl}}
\author[M. S. Petersen, M. Roule, J.-B. Fouvry, C. Pichon, and K. Tep]{Michael S. Petersen$^{1,2}$, Mathieu Roule$^{2}$, Jean-Baptiste Fouvry$^{2}$, Christophe Pichon$^{2,3,4}$ \newauthor and Kerwann Tep$^{2}$
\vspace*{6pt}\\
\noindent $^{1}$Institute for Astronomy, University of Edinburgh, Royal Observatory, Blackford Hill, Edinburgh EH9 3HJ, UK\\
\noindent $^{2}$Institut d'Astrophysique de Paris, UMR 7095, 98 bis Boulevard Arago, F-75014 Paris, France\\
\noindent $^{3}$IPhT, DRF-INP, UMR 3680, CEA, L’orme des Merisiers, Bât 774, 91191 Gif-sur-Yvette, France\\
\noindent $^{4}$Korea Institute for Advanced Study, 85 Hoegi-ro, Dongdaemun-gu, Seoul 02455, Republic of Korea
}
\maketitle

\begin{abstract}
We present \linearresponse\@, an efficient, versatile public  library written in \julia\ to compute the linear response of self-gravitating ($3D$ spherically symmetric) stellar spheres and ($2D$ axisymmetric razor-thin) discs. 
\linearresponse\ can scan the whole complex frequency plane, probing unstable, neutral and (weakly) damped modes.
Given a potential model and a  distribution function, this numerical toolbox estimates the modal frequencies as well as the shapes of individual modes. 
The libraries are validated against a combination of previous results for the spherical isochrone model and Mestel discs, and new simulations for the spherical Plummer model.
Beyond linear response theory, the realm of applications of \linearresponse\ also extends to the kinetic theory of self-gravitating systems through a modular interface.
\end{abstract} 

\begin{keywords} Gravitation -- Instabilities -- Galaxies: kinematics and dynamics -- Software: public release\end{keywords}

%%%%%%%%%%%%%%%%%%%%
\section{Introduction}
\label{sec:introduction}
%%%%%%%%%%%%%%%%%%%%
The response of self-gravitating systems to perturbations has been studied for more than half a century, mostly using the {\it matrix method} \citep{Kalnajs.1971}.
As discussed in detail in section~{5.3.2} of \citet{Binney.Tremaine.2008}, this method uses basis functions to represent the gravitational response of the stellar system.
The response matrix $\bM$ at a given complex frequency $\omega$ is generically
\begin{equation}
    \label{eq:ResponseMatrix}
    \bM (\omega) = \sum_{\bn}\! \int\limits_{\cal L} \!\!\rd\bJ\frac{\bG_{\bn}(\bJ)}{\bn\!\cdot\!\bOmega\left(\bJ\right)-\omega}.
\end{equation}
This equation encodes the physics of (linear) stability in self-gravitating systems.
Here, $\sum_{\bf n}$ is the sum over all (allowed) resonance vectors ${\bn}$; $\int_{\cal L} d{\bf J}$ is the scan over the populated orbital (action) space (with an appropriate Landau prescription); ${\bn\!\cdot\!\bOmega\left(\bJ\right)\!-\!\omega}$ is the resonant amplification at a given (complex) frequency $\omega$ and for orbital frequencies $\bOmega$; ${\bG_{\bn}}$ are functions of the system's \DF\ and are matrices by virtue of decomposing the Newtonian pairwise interaction on potential basis elements that encode the long-range nature of the gravitational interaction~\citep{Kalnajs.1976}. 

The system sustains a mode at a complex frequency $\omega$ when the condition
\begin{equation}
    \label{eq:def_dielectric}
    \det[\bvareps(\omega)]=0, \quad \mathrm{with} \; \bvareps(\omega) = \bI - \bM(\omega),
\end{equation} 
is met, where $\bI$ is the identity matrix and $\bvareps$ is called the (gravitational) dielectric matrix. 
We write frequencies satisfying this equality as $\omegaM$, which we further break down into ${\omegaM\!=\! \OmegaM \!+\!\ri\gammaM }$, with the oscillation frequency, $\OmegaM$, and the growth rate, $\gammaM$.
Modes may appear at any point in the complex plane. Modes with ${\gammaM\!>\!0}$ are unstable, modes with ${\gammaM \!=\!0}$ are neutral, and modes with ${\gammaM \!<\!0}$ are damped. 

Previous studies in linear response have probed modal structure in self-gravitating discs \citep[e.g.,][]{Zang.1976,Toomre.1981,Vauterin.Dejonghe.1996,Pichon.Canon.1997,Evans.Read.1998.I,Jalali.Hunter.2005,Fouvry.etal.2015,deRijcke.Voulis.2016} and spheres \citep[e.g.,][]{Polyachenko.Shukhman.1981,Weinberg.1989,Weinberg.1991,Saha.1991,Bertin.1994,Murali.Tremaine.1998,Rozier.etal.2019,Fouvry.Prunet.2022,Weinberg.2023}. 
Most calculations have been interested in studying instabilities, i.e., modes with ${\gammaM \!>\! 0}$ \citep[for a detailed look at stability, see][]{Palmer.1994}. 
Traditionally, finding damped modes has been challenging because of  the analytical continuation required by Landau's prescription.
\cite{Zang.1976} presented  a first prediction in razor-thin discs
while relying heavily on the scale invariance of the Mestel potential.
\citet{Weinberg.1994} broke through by using rational functions to (numerically) continue calculations made in the upper half frequency plane to the lower half, finding a lopsided damped mode in the King family of spherical models.
Standard techniques in plasma physics offer other options for analytic continuation, which were applied to stellar systems by~\citet{Fouvry.Prunet.2022}. 
Damped modes are interesting for their capacity to dissipate energy through interactions of the mode and resonating stellar orbits \citep{Nelson.Tremaine.1999}.
This phenomenon may have implications for real astrophysical systems, including accelerating the overall relaxation \citep[see, e.g.,][]{Hamilton.etal.2018, Heggie.Breen.Varri.2020}, as well as the secular dynamics captured by the Balescu--Lenard equation \citep{Heyvaerts.2010,Chavanis.2012}, and the so-called mode-particle interactions~\citep{Hamilton.Heinemann.2020}.

In this paper, building upon the method developed in~\citet{Fouvry.Prunet.2022}, hereafter \citetalias{Fouvry.Prunet.2022}, we introduce software libraries in the \julia\ language to perform linear response calculations.
Details of the toolbox are given in Appendix~\ref{app:Software}.
Linear response calculations are intricate, requiring expensive integrals over the full action phase that must be performed with care. 
In the past, codes and methods for linear response have often not been publicly shared, necessitating complicated re-implementations. 
By publishing our code, we hope to create a framework where others can build on the methods, or add their own. 
By using \julia\@, we are able to rapidly explore parameter space and convergence rates, while retaining a high level of readability and optional interactivity.

The paper is organised as follows.
In Section~\ref{sec:linear}, we sketch the linear response computations and present the \julia\ libraries developed to tackle them.
In Section~\ref{sec:Plummer}, we compute the response for the spherical Plummer model, probing its modal spectrum across a range of \DFs\@, and compare with $N$-body simulations.
In Section~\ref{sec:Discs}, we validate our extension to the linear response of razor-thin discs on the well-studied constant circular velocity discs~\citep{Zang.1976}.
We wrap up and conclude in Section~\ref{sec:Conclusions}.
Throughout the main text, technical details are kept to a minimum and deferred to appendices or to relevant references.

%%%%%%%%%%%%%%%%%%%%
\section{Linear response of spheres and discs}
\label{sec:linear}
%%%%%%%%%%%%%%%%%%%%

The linear response of spheres and discs can be split in independent harmonics $\ell$ (historically denoted $m$ for the discs).
For these systems and within the appropriate ``resonance'' coordinate system $(u,v)$ introduced in equation~{(10)} of \citetalias{Fouvry.Prunet.2022}, each element of the response matrix of harmonic $\ell$ reads 
\begin{equation}
    \label{eq:M_int_u}
    M^{\ell}_{pq}(\omega) = \sum_{\bn\in\mathbb{Z}^2} \intLb \!\! \rd u \frac{G^{\ell\bn}_{pq} (u)}{u - \varpi_{\bn} (\omega)} ,
\end{equation}
where the functions\footnote{Note that $G^{\ell \bn}_{pq} (u)$ in equation~\eqref{eq:M_int_u} slightly differs from $\bG_{\bn}(\bJ[u])$ in equation~\eqref{eq:ResponseMatrix}. See Appendix~\ref{app:ComputingG} for details.} $G^{\ell\bn}_{pq}(u)$ depend on the Fourier transform of bi-orthogonal basis elements (indexed by $p$ and $q$), the resonance number, ${\bn\!=\!(n_1,n_2)}$, and involve an integral over some other coordinate ${ \rd v }$, as explained in Appendix~\ref{app:ComputingG}. 
These functions, which depend on the dimensionality of the problem, are detailed in Appendix~\ref{app:LRIntegrands}.
In equation~\eqref{eq:M_int_u}, ${\varpi_{\bn} (\omega)}$ stands for the rescaled (complex) frequency introduced in equation~{(11)} of \citetalias{Fouvry.Prunet.2022}.
Importantly, the imaginary parts of the rescaled frequency $\varpi_{\bn}$ and ${\omega}$ share the same sign.
Computing the linear response of a stellar system mostly entails carrying out the integrals outlined in equation~\eqref{eq:M_int_u}. It is crucial to focus on the resonant denominator, adhering to Landau's prescription as detailed in, for instance, equation~{(2)} in the work by \citetalias{Fouvry.Prunet.2022}.
The goal behind producing \linearresponse\ is to extend the prescription used for the isochrone sphere in~\citetalias{Fouvry.Prunet.2022} to any numerically-given potential and \DF\@, as well as to discs. 
To be as versatile as possible, the software is decomposed in four libraries.
\begin{itemize}
    \item \orbitalelements\ (Appendix~\ref{app:ComputingFrequencies}), which, given any central potential and its two first derivatives, performs various changes of orbital coordinates to compute the actions $(\Jr,L)$, orbital frequencies $(\Omega_1,\Omega_2)$ and resonance coordinates $(u,v)$.\footnote{At present, the mappings are restricted to cored potentials. Extending them to cuspy potentials will be the purpose of future work.}  
    \item \astrobasis\ (Appendix~\ref{app:ComputingBasis}) provides bi-orthogonal bases of potential-density pairs for spheres and discs. This library is constructed such that a user could readily supply their own additional bases using a straightforward template.
    \item \fht\ (Appendix~\ref{app:LegendreIntegration}) performs the (Landau's prescription-compliant) finite Hilbert transform using Legendre polynomials, as detailed in appendix~{D} of \citetalias{Fouvry.Prunet.2022}.
    \item \linearresponse\ (Appendix~\ref{app:ComputingLR}) is the driver library. Relying on the three (independent) previous libraries, \linearresponse\ implements the computation of the response matrix ${\bM^{\ell}\!(\omega)}$. It performs the Fourier transform of the basis elements, then computes the functions $G^{\ell\bn}_{pq}(\bJ)$, before finally computing the finite Hilbert transform.
\end{itemize}

We refer to the appendices for details on the implementation and to the respective GitHub repositories for installation and usage.
The bulk of the following sections is devoted to exercising the library in a series of simple systems.

%%%%%%%%%%%%%%%%%%%%
\section{Plummer sphere: theory and simulations} 
\label{sec:Plummer}
%%%%%%%%%%%%%%%%%%%%

In order to test and validate \linearresponse\@, we first examine models representing spherically symmetric globular clusters.
Following~\citetalias{Fouvry.Prunet.2022}, we consider the isochrone potential in Appendix~\ref{app:Isochrone}.
We use its analytical expressions to check our numerical calculation of the orbital elements, and recover all of \citetalias{Fouvry.Prunet.2022}'s results.
This is reassuring.  We also point out two caveats raised by our exploration, namely (i)  diverging gradients of the anisotropic \DF\ at the edge of the domain leading to convergence issues in the ${\ell\!=\!2}$ case, and (ii) a (neutral) translation mode that becomes damped in the ${\ell\!=\!1}$ case.

In this section, we expand our linear response study to the \citet{Plummer.1911} model. In particular, we first study the (${\ell\!=\!2}$) \ROI\@ for radially anisotropic models (Section~\ref{sec:PlummerROI}), before investigating isotropic \DFs\ and ${\ell\!=\!1}$ modes (Section~\ref{sec:PlummerISO}). 
We then simulate the Plummer model with \EXP\@, a basis function expansion $N$-body code, to test our predictions (Section~\ref{sec:SimulationResults}).
We provide the relevant equations for the Plummer model in Appendix~\ref{app:Plummer}.

%%%%%%%%%%%%%%%%%%%%
\subsection[l = 2 modes -- Radial Orbit Instability]{${\ell\!=\!2}$ modes -- Radial Orbit Instability} 
\label{sec:PlummerROI}
%%%%%%%%%%%%%%%%%%%%

The \ROI\@ arises in models with some degree of radial anisotropy, i.e., the \DF\ contains an enhancement at low angular momentum values \citep[for a review, see][]{Marechal.Perez.2011}. 
Often, this is accomplished via the introduction of some anisotropy radius, $\rra$, outside of which the fraction of radial orbits is enhanced. 
The mechanisms for the instability are discussed in \citet{Palmer.1994,Polyachenko.Shukhman.2015}. 
The classic \ROI\@ results in a quadrupole (${\ell\!=\!2}$) zero oscillation frequency mode (${ \OmegaM \!=\! 0 }$) that grows exponentially in time (${\gammaM\!>\!0}$). 
In practice, the \ROI\@ is generated by the \ILR\@, $\bn = (-1,2)$, and its opposite. 
This simplicity makes it a particularly compelling instability to use as benchmark.
%Owing to this, only in this section, we simplify the sum over resonances in equation~\eqref{eq:ResponseMatrix} to this dominating resonance. \MR{Is it true or are you using fiducial \texttt{n1max} finally?}

Various studies have sought to establish the boundary for stability in the radially anisotropic Plummer model. \cite{Dejonghe.Merritt.1988} estimated the threshold for stability in the Plummer model to be ${\rra/\brc\!\simeq\! 0.9}$ using the criteria of \citet{Barnes.Goodman.Hut.1986},
with $\brc$ the Plummer's scale radius (equation~\ref{eq:Plummer_Pot}).
Using the mean-field code from \citet{Merritt.1987},
\cite{Dejonghe.Merritt.1988} also reported that the stability threshold was somewhere around ${\rra/\brc\!\simeq\!1.1}$.
Relying on direct $N$-body simulations, \citet{Breen.Varri.Heggie.2017}
confirmed that ${\rra/\brc\!=\!0.75}$ was unstable,
but, in contrast to \cite{Dejonghe.Merritt.1988}, \citet{Breen.Varri.Heggie.2017} found that ${\rra/\brc\!=\!1.0}$ was stable.

\begin{figure*}
    \begin{center}
        \includegraphics[width=1.0\textwidth]{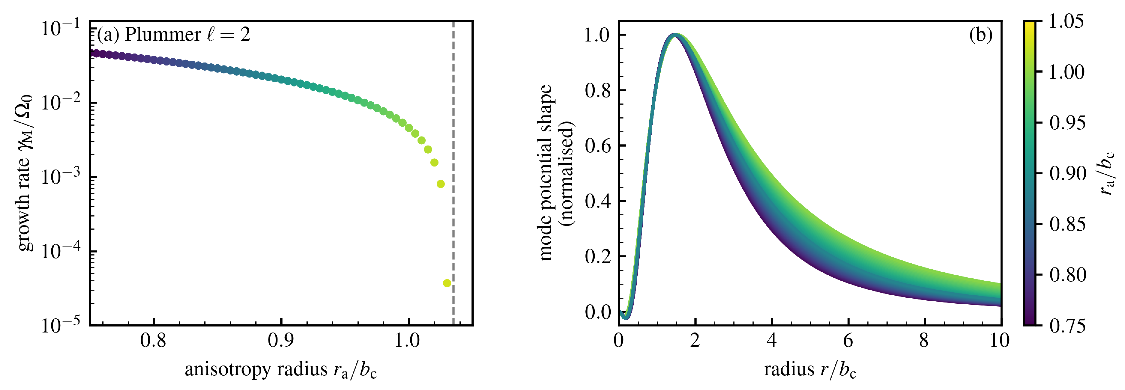}
        \caption{\label{fig:Figure1} 
        \textit{Left panel}: Growth rate ${\gammaM\!=\!\ImPart[\omegaM]}$ of the radial orbit instability (ROI) as a function of the anisotropy radius $\rra$ for the Plummer model. 
        \textit{Right panel}: Predicted potential fluctuation as a function of radius for the \ROI\@-unstable modes in the Plummer model. 
        Different curves correspond to different values of $\rra$, which are colour-coded. 
        Even though the growth rate drops by three orders of magnitude, the shape of the mode changes very little.
        All calculations are made using the Fiducial settings.
        }
    \end{center}
\end{figure*}

We test our linear response machinery on a range of $\rra/\brc$ values.
Fig.~\ref{fig:Figure1} shows the results of searching for modes using a ``Fiducial'' set of control parameters.\footnote{The ``Fiducial'' control parameters for spherical calculations take the software defaults for \orbitalelements\ (Table~\ref{tab:OEparams}) and \linearresponse\ (Table~\ref{tab:LRparams}), along with a basis scale ${\rrb\!=\!5.0}$ and 100 radial basis functions. For ${\ell\!=\!2}$ calculations, this results in 62 resonances to calculate. For ${\ell\!=\!1}$ calculations, this results in 42 resonances to calculate. For the Plummer sphere, the fiducial value of $\Omega_0$ is 2, cf. equation~\ref{eq:plummeromega0}.}
We find a threshold for instability of ${\rra/\brc \!=\!1.035}$, as highlighted by the vertical dashed line in Fig.~\ref{fig:Figure1}.
Let us finally note that, here, the response matrix was computed using the basis provided by \citet{CluttonBrock.1973} tailored for the Plummer potential.
Hence, the basis matches the underlying mean profile at large radii: this drastically reduces numerical noise. 
We report results for varying control parameters in Table~\ref{tab:ROIResults}, finding that the uncertainty in growth rate from control parameters is a few percent.
This leads to no alteration in the threshold for instability.
%%%%%%%%%%%%%%%%%%%%
%%%%%%%%%%%%%%%%%%%%
%
Since individual $\ell$ harmonics are decoupled in linear response calculations, we investigate ${\ell\!=\!(1,3,4)}$ for similar instability behaviour. We do not find evidence for any modes for ${\ell\!\ne\!2}$.

%%%%%%%%%%%%%%%%%%%%
\subsection[A search for l = 1 damped modes]{A search for ${\ell\!=\!1}$ damped modes}
\label{sec:PlummerISO}
%%%%%%%%%%%%%%%%%%%%

We next search for evidence of damped ${\ell\!=\!1}$ modes, akin to the findings of \citet{Weinberg.1994} and \citet{Heggie.Breen.Varri.2020} for King models. 
Using an isotropic \DF\ (see Appendix~\ref{app:Plummer}), we perform a search in the complex plane. 
We find zeros in the determinant of the dielectric matrix $\bvareps$, but these appear to be spurious non-converged realisations of the neutral translation mode. 
The right column of Table~\ref{tab:ISOResults} lists the results for different configurations. 
As the number of resonances considered increases, the predicted mode's frequency shifts towards the origin of the complex frequency plane, as in the isochrone model (Appendix~\ref{app:IsochroneISO}). 
In addition, the ``false'' mode clearly resembles the density derivative ${\rd\rho/\rd r}$, i.e., the modal shape associated with an infinitesimal translation of the cluster.
As these unconverged poles are the only one we found, we do not predict any robust ${\ell\!=\!1}$ modes.
For completeness, we also search other harmonics ${\ell\!=\!(2,3,4)}$, finding no evidence for damped modes.

\begin{figure}
    \begin{center}
        \includegraphics[width=0.45\textwidth]{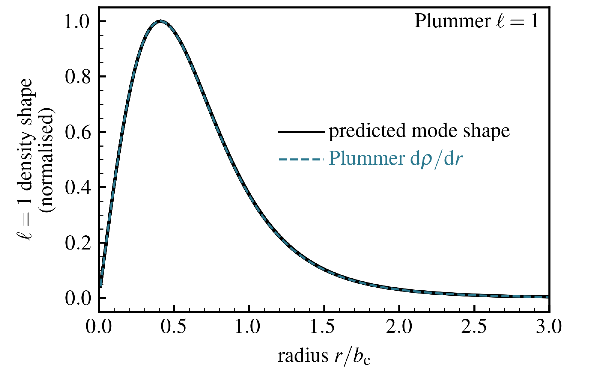}
        \caption{\label{fig:Figure2} 
        Predicted density mode shape vs.\ radius for ${\ell\!=\!1}$ damped modes, using several combinations of configuration parameters. 
        Predictions from different configurations are indistinguishable in mode shape, despite a range predictions for $\omegaM$ (see Table~\ref{tab:ISOResults}). 
        Predictions for the Plummer model are shown in black. 
        The dashed curve is ${\rd\rho/\rd r}$ for the Plummer model. 
        The predicted mode shapes are indistinguishable from ${\rd\rho/\rd r}$.}
    \end{center}
\end{figure}

Fig.~\ref{fig:Figure2} shows the predicted mode shape with radius for the Plummer model (black curve), as compared to ${\rd\rho/\rd r}$. 
As expected given the results of~\citetalias{Fouvry.Prunet.2022}, the predicted mode appears to be the neutral ${\omegaM\!=\!0\!+\!0\ri}$ mode, pushed slightly away from the origin of the complex frequency plane due to incomplete convergence. 
We show the mode shape for the full set of configurations (Fig.~\ref{fig:Figure2}), but all curves lie on top of one another. 
Surprisingly, even though the frequency of the mode is not converged, its shape appears to be.
Here, we recover again that reconstructing a large-scale translation mode from orbital space integrals and resonant interactions is, by design, a challenging task.
%%%%%%%%%%%%%%%%%%%%
%%%%%%%%%%%%%%%%%%%%

%%%%%%%%%%%%%%%%%%%%
\subsection[N-Body Simulation]{$N$-Body Simulation}
\label{sec:SimulationResults}
%%%%%%%%%%%%%%%%%%%%

In this section, we report the results of $N$-body simulations of the Plummer model. 
We test two different \DFs\ (see Appendix~\ref{app:Plummer}): one radially anisotropic, and one isotropic. 
We realise the $N$-body simulations using initial conditions from \texttt{PlummerPlus} \citep{Breen.Varri.Heggie.2017}. 
We run the models using \EXP\ \citep{Weinberg.1999,Petersen.Weinberg.Katz.2022}, a basis function expansion $N$-body code. 
Further numerical details for both the initial conditions and the $N$-body integration are given in Appendix~\ref{app:NBody}.

%%%%%%%%%%%%%%%%%%%%
\subsubsection{Anisotropic simulations, Radial Orbit Instability}
\label{sec:PlummerROImeasurements}
%%%%%%%%%%%%%%%%%%%%

To attempt to recover the \ROI\@ in the Plummer clusters, we select six different values of $\rra / \brc$ for which to run $N$-body simulations: ${[0.75,0.80,0.85,0.90,0.95,1.0]}$. 
These values span the entire range of models predicted to be unstable. 
For each $\rra/\brc$, we run an ensemble of six simulations of $2^{18}$ particles each.

Fig.~\ref{fig:Figure3} shows the results for the ${\rra/\brc\!=\![0.75,0.80,0.85,0.9]}$ runs using \EXP\ (see Appendix~\ref{app:NBody}). 
In general, the runs for any value of $\rra/\brc$ show similar behaviour, though the dispersion between realisations increases with $\rra/\brc$ (i.e., dispersion between realisations increases as the model approaches marginal stability). 
The solid curves in the right panel of Fig.~\ref{fig:Figure3} show the reconstruction of the potential fluctuations for the series of $\rra/\brc$ models; the qualitative agreement with the predicted mode shapes and trends (shown as dashed curves) with $\rra/\brc$ are remarkably high, though the $N$-body peak is consistently 10 per cent lower than the predicted peak. 
Measurements of the mode shape during the growth are consistent with the same shape.
In the runs with ${\rra/\brc\!=\!0.95}$, only three of the six runs performed appear to show any instability. 
Thus, ${\rra/\brc\!=\!0.95}$ appears to be a weakly unstable model.
We do not measure an instability for any of the ${\rra/\brc\!=\!1.0}$ runs, in $N$-body simulations.
%%%%%%%%%%%%%%%%%%%%
\begin{figure*}
    \begin{center}
        \includegraphics[width=1.0\textwidth]{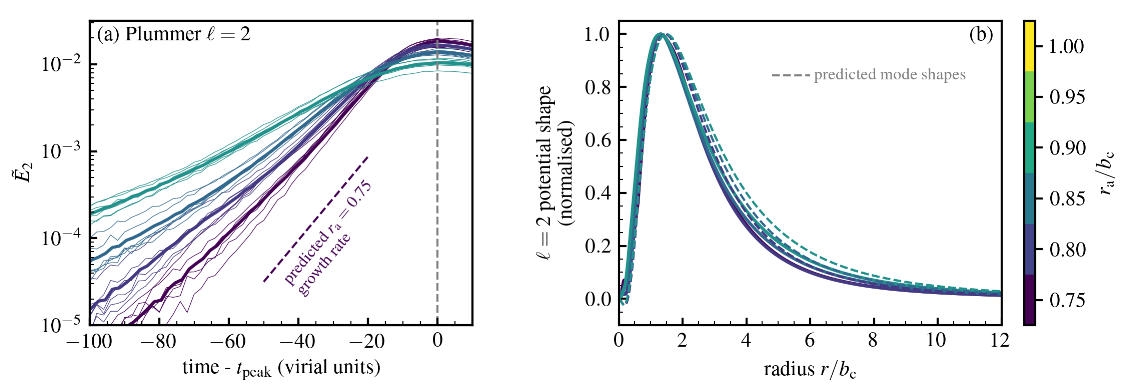}
        \caption{\label{fig:Figure3} 
        \textit{Left panel}: Growth of the ${\ell\!=\!2}$ potential fluctuations energy ($\tE_{\ell}$, equation~\ref{eq:def_tE}),
        through time in $N$-body simulations of the radially-anisotropic Plummer sphere for four different anisotropy radii, $\rra/\brc$. Simulations with $\rra/\brc=0.95$ and $\rra/\brc=1.00$ are not shown, as instabilities were not reliably measured.
        Thin curves are individual simulation realisations; thick curves are the ensemble average for each value of $\rra/\brc$. 
        Curves are aligned on the time of the first peak of the potential perturbation, denoted ${t\!=\!0}$, where
        the shape of each mode is measured. 
        The exponential growth predicted by linear theory is clearly visible before the modes' saturation. 
        \textit{Right panel}: Shape of the ${\ell\!=\!2}$ potential fluctuations measured from the $N$-body simulations (solid curves). 
        The dashed curves are the corresponding predictions from \linearresponse\@.
        The linear predictions and numerical measurements are in pleasant agreement.
        }
    \end{center}
\end{figure*}
%%%%%%%%%%%%%%%%%%%%

Table~\ref{tab:Table1} compares the measured growth rates from the $N$-body simulations to the predictions from \linearresponse\@. 
The agreement is quantitatively good, in that all measured growth rates are within ${3\sigma}$ of the predicted growth rate, and both approaches recover the qualitative trend of decreasing $\gammaM$ with increasing $\rra/\brc$. 
We do not find evidence for a second unstable mode in the simulations. 
The numerical values from the Fiducial run, except with a basis scalelength of ${\rrb\!=\!1}$, match the basis used to run the simulation parameters (Appendix~\ref{app:NBody}).
This ensures a straightforward comparison with the $N$-body simulations.
We find that the predicted eigenvectors (equation~\ref{eq:def_Xp}) qualitatively match the basis function coefficients in the $N$-body simulations.
%%%%%%%%%%%%%%%%%%%%
\input{tables/Table1}
%%%%%%%%%%%%%%%%%%%%

In conclusion, the $N$-body simulations and the predicted linear response growth rate and mode shape match to a high degree, validating both our linear response approach, and the ability of \EXP\ to make intricate dynamical measurements.
Future work can increase the number of particles in the runs to validate the predicted growth rates, and attempt to recover lower growth rates near the stability boundary.

%%%%%%%%%%%%%%%%%%%%
\subsubsection{Isotropic simulations, ${\ell\!=\!1}$ measurements}
\label{sec:PlummerISOmeasurements}
%%%%%%%%%%%%%%%%%%%%

By analogy with the investigation of Section~\ref{sec:PlummerISO}, we perform an ensemble of six simulations of Plummer spheres with an isotropic \DF\ (see Appendix~\ref{app:NBody}).
The left panel of Fig.~\ref{fig:Figure4} shows the run with time of the (rescaled) energy, ${\tilde{E}_{1}\!\propto\!E_{\ell=1} }$ (Eq.~\ref{eq:def_tE}), for each simulation.
Interestingly, $\tilde{E}_{1}$ increases and slowly approaches a saturated level that differs between realisations.
Upon inspection, the amplitude $\tilde{E}_{1}$ is dominated by the lowest-order function in each realisation, i.e., these fluctuations are large-scale.
None of the basis elements show clear evidence of a coherent oscillation frequency.
%%%%%%%%%%%%%%%%%%%%
\begin{figure*}
    \begin{center}
        \includegraphics[width=1.00\textwidth]{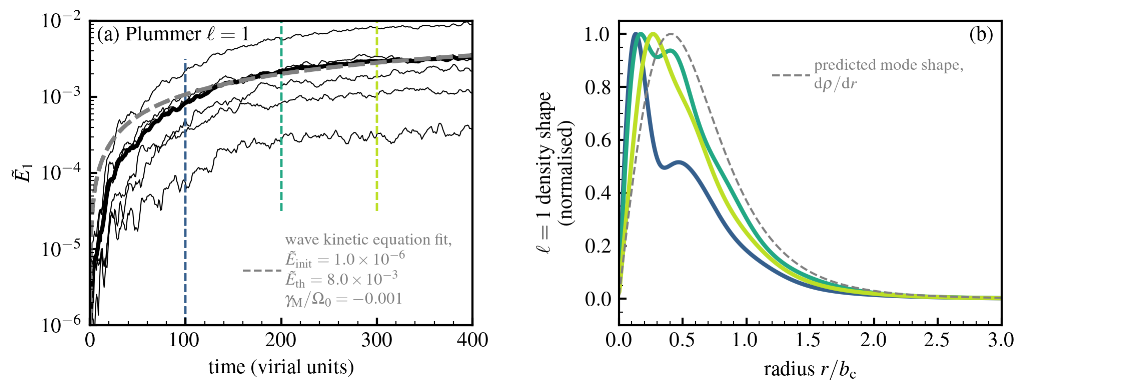}
        \caption{\label{fig:Figure4} 
        \textit{Left panel}: Amplitude of the ${\ell\!=\!1}$ potential fluctuations energy ($\tE_{\ell}$, equation~\ref{eq:def_tE}) vs.\ time in the isotropic $N$-body realisations of Plummer spheres.
        Surprisingly, this time evolution could be compatible with a wave kinetic equation, as in equation~\eqref{eq:wave_kinetic}. 
        The best-fit solution is shown as a dashed gray curve. 
        \textit{Right panel}: Radial shape of ${\ell\!=\!1}$ density fluctuations measured from the $N$-body simulations at three different times (denoted by vertical coloured dashed lines in the left panel). 
        The dashed grey curve in the right panel shows ${\rd\rho/\rd r}$ vs.\ radius for the Plummer cluster.
        As time increases, the measured dipolar perturbations get more and more similar to a simple translation of the cluster as a whole.
        }
    \end{center}
\end{figure*}
%%%%%%%%%%%%%%%%%%%%

The right panel of Fig.~\ref{fig:Figure4} may be compared with Fig.~\ref{fig:Figure2} for the predictions of modal shapes.
While the linear predictions are consistent with the only apparent mode being the neutral mode with shape ${\rd\rho/\rd r}$, the results from the $N$-body simulation are generally not consistent with ${\rd\rho/\rd r}$. 
Instead, the peak density deviation is confined to smaller radii. 
Over time in the simulation, the lowest-order function dominates further, such that the shape of the ${\ell\!=\!1}$ density fluctuations further resembles ${\rd\rho/\rd r}$.
However, over the duration of the simulations here, the cluster does not fully offset.
Rather, the inner regions of the cluster offset, while the outer regions appear to remain roughly fixed. 
This likely owes to the long dynamical times in the cluster's outskirts. 

Interestingly, in Fig.~\ref{fig:Figure4}, the time evolution of the total energy in the ${\ell\!=\!1}$ fluctuations shows a clear sign of saturation.
\cite{Rogister.Oberman.1968}, equation~{(21)} therein, put forward a wave kinetic equation~\citep[see also equation~{22} in][]{Hamilton.Heinemann.2020}
that describes the (linear) saturation of potential fluctuations in self-gravitating systems.
It reads
\begin{equation}
    \label{eq:wave_kinetic}
    \frac{\rd \tE_{\ell}}{\rd t} = 2\gammaM \left(\tE_{\ell} - \tEth \right),
\end{equation} 
where ${\tE_{\ell}}$ is the (rescaled) energy in the ${ \ell \!=\! 1 }$ gravitational fluctuations and $\tEth$ the (rescaled) asymptotic level of dressed fluctuations.
Equation~\eqref{eq:wave_kinetic} states that in a self-gravitating system which sustains a (weakly) damped mode with damping frequency $\gammaM$, transients only fade away after a (few) ${ 1 / \gammaM }$, at which stage fluctuations are fully dressed by self-gravity~\citep[see, e.g.,][]{Hamilton.Heinemann.2023}.
Solutions of equation~\eqref{eq:wave_kinetic} are readily obtained. 
Given the run of $\tE_{\ell=1}$ in Fig.~\ref{fig:Figure4}, we can use the ensemble average ${\ell\!=\!1}$ energy (the thick black curve in the left panel of Fig.~\ref{fig:Figure4}) to find best-fitting values for $\tEth$, $\gammaM$ and the initial energy fluctuations, ${\tEinit \!=\! \tE_{\ell} (t\!=\!0)}$. 
We find ${\tEinit \!=\!1.0\!\times\!10^{-6}}$, ${\tEth\!=\!8.0\!\times\!10^{-3}}$, and ${\gammaM/\Omega_{0}\!=\!-0.001}$. 
The time evolution of the numerical simulations
appears to be reasonably compatible with the wave kinetic equation in the isotropic case. 
However, the fitted solution for $\gammaM$ is outside of the region that is numerically accessible with the current methods used in \linearresponse\@, despite implying a relatively long time for the ${\ell\!=\!1}$ energy to reach a thermal state. 
Additionally, the apparent scatter of values of $\tEth$ for individual runs
is somewhat unexpected. 
We defer further investigation to future work.

%%%%%%%%%%%%%%%%%%%%
\section{Mestel razor-thin discs: validation}
\label{sec:Discs}
%%%%%%%%%%%%%%%%%%%%

To emphasize the generality of \linearresponse\@, we now consider  razor-thin (${2D}$) discs.
In that case, stars still interact through the usual Newtonian interaction potential ${U(\br,\brp)\!=\!-G/|\br-\brp|}$, with $G$ the gravitational constant, but are confined to a plane.
Assuming an axisymmetric mean \DF\@, the stars' mean-field orbits can be described by the two action variables $(\Jr,L)$, just like in the case of spheres.
Hence, all the previous numerical methods from \citetalias{Fouvry.Prunet.2022} naturally apply to razor-thin discs.
The only required modifications are: 
(i) marginally modifying the linear response integrand, ${G^{\ell\bn}_{pq}}$, in equation~\eqref{eq:M_int_u}, given in Appendix~\ref{app:G_discs}, and
(ii) implementing ${2D}$ bi-orthogonal basis elements \citep[e.g.,][]{CluttonBrock.1972,Kalnajs.1976}, given in Appendix~\ref{app:2Dbasis}.

In order to validate our implementation, we aim to recover well-documented unstable modes in razor-thin discs.
In practice, following the work of \cite{Zang.1976}, we consider Mestel discs \citep{Mestel.1963}, whose scale-invariance allows for various analytical simplifications.
Nonetheless, in what follows, we do not use these simplifications and rather use the generic scheme from \linearresponse\@.
Following \cite{Evans.Read.1998.I}, the \DF\ of the stars is tapered in the central region.
The presence of an inner cut-off mimicks an unresponsive central bulge, hence introducing a reflexive boundary\footnote{The Mestel potential diverges for ${ r\!\to\!0 }$, making exactly radial orbits singular. As a result, we also softened the mean gravitational potential in the central region (Appendix~\ref{app:DiscsDF}). This softening is on a much smaller scale than the \DF\@'s inner cut-out, and does not impact the linear response predictions.}: the sharper this inner cut-out, the stronger the instability \citep{Zang.1976}.
The disc's outskirts are also tapered, though this does not impact the disc's stability, provided that this external cut-out is sharp enough and far enough \citep{Evans.Read.1998.II}.
We refer to Appendix~\ref{app:DiscsDF} for  details and notably the definition of the scaling frequency $\Omega_0$.
Note that the method based on finite Hilbert transform implemented in \linearresponse\ is not perfectly suited here.
Indeed, because of its central divergence, the frequency range in Mestel discs is not finite. 
We deal with this particular issue by limiting the orbital domain probed by the code, constrained by the inner \DF\ taper (see Appendix~\ref{app:ResonancesVariables} for details).
It would be rewarding to extend \citetalias{Fouvry.Prunet.2022}'s methodology to the case of semi-finite frequency supports.

Once this model is set up, we perform stability analysis for two-armed modes, i.e., ${\ell\!=\!2}$ modes, as one varies the properties of the inner taper.\footnote{The azimuthal harmonic number for discs is historically denoted ${m}$. Adapting here from the spherical case, we nonetheless denote it $\ell$.}
As reported in Table~\ref{tab:ZangModes}, we find a satisfying agreement between the (semi-analytical) predictions of~\cite{Zang.1976} and~\cite{Evans.Read.1998.II}, and the present linear predictions for the growth rate and oscillation frequency of the most unstable mode. 
These predictions have already been confirmed using numerical simulations by \cite{Sellwood.Evans.2001}.
%%%%%%%%%%%%%%%%%%%%
\input{tables/Table2}
%%%%%%%%%%%%%%%%%%%%

\begin{figure}
    \begin{center}
        \includegraphics[width=0.5\textwidth]{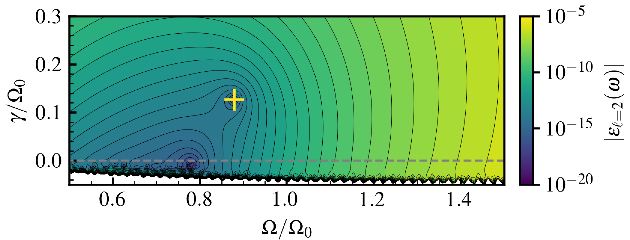}
        \caption{\label{fig:Figure5} 
        Isocontours of the determinant of the ${\ell\!=\!2}$ (gravitational) dielectric matrix from equation~\eqref{eq:def_dielectric} for the ${N\!=\!4}$ Zang disc.
        The dominant mode obtained by~\citet{Zang.1976} is highlighted 
        with a yellow cross and is recovered within $1\%$ precision.}
    \end{center}
\end{figure}

\begin{figure}
    \begin{center}
    \includegraphics[width=0.45\textwidth]{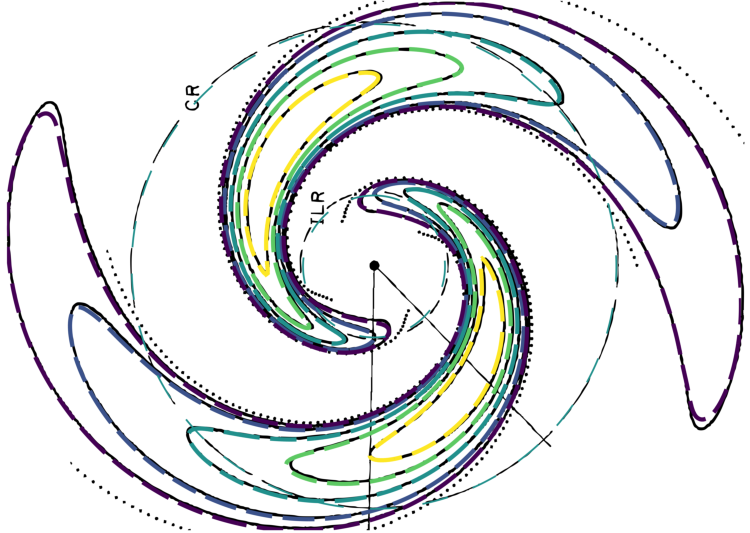}
        \caption{\label{fig:Figure6} 
        Shape of the ${\ell\!=\!2}$ dominant mode of the ${N\!=\!4}$ Zang disc as predicted by \linearresponse\ (coloured dashed lines) overlaid with the shape obtained by \protect\cite{Zang.1976} (in black, figure 9 therein).
        For both shapes, the contours denote the $10$, $20$, $40$, $60$ and $80\%$ of the peak density, with only the overdensity being represented.
        The dotted circles show the corotation (CR) and inner Lindblad resonance (ILR) radii of the mode.
        Both linear predictions are in very good agreement.}
    \end{center}
\end{figure}

In Fig.~\ref{fig:Figure5}, we present a typical map of the complex frequency plane one can obtain from \linearresponse.
More precisely, Fig.~\ref{fig:Figure5} illustrates the determinant of the (gravitational) dielectric matrix, ${ \bvareps (\omega) }$ from equation~\eqref{eq:def_dielectric}, through its level contours in the complex-frequency plane.
This determinant vanishes at the frequency ${\omegaM/\Omega_0\!=\!0.878\!+\!0.126\ri}$, i.e., the system supports a growing mode.
The shape of this unstable mode is reported in Fig.~\ref{fig:Figure6}, where we compare it with the result from \cite{Zang.1976}.
We find a quantitative match between both approaches.
In Fig.~\ref{fig:Figure5}, the saturation and ringing for damped frequencies are to be expected. They are a direct consequence of analytical continuation being an ill-conditioned numerical problem.  
Further explanation is given in Appendix~\ref{app:LegendreIntegration}.
%%%%%%%%%%%%%%%%%%%%
%%%%%%%%%%%%%%%%%%%%
%%%%%%%%%%%%%%%%%%%%
%%%%%%%%%%%%%%%%%%%%

As a concluding remark, let us note that here we used the generic basis from~\cite{CluttonBrock.1972}, which is not tailored to asymptotically match the disc's
underlying potential. 
Interestingly, this did not impact our ability to recover precisely the underlying modes.
Expanding \linearresponse\ to accommodate for more generic basis elements will nonetheless be the topic of future work.

%%%%%%%%%%%%%%%%%%%%
\section{Summary and Conclusions}
\label{sec:Conclusions}
%%%%%%%%%%%%%%%%%%%%

In this paper, we presented \linearresponse\@, an efficient, readable software written in \julia\ to perform linear response calculations for self-gravitating stellar spheres and discs. 
Given a model's potential and \DF\@, it can predict the system's modal response, as well as the shapes of individual modes. 
We validate the four underlying libraries against known results for the isochrone sphere, before investigating the Plummer model.

In the case of the isochrone model, we find quantitative agreement with previous work.
Studying the ${\ell\!=\!1}$ response, we confirm that the results of~\citetalias{Fouvry.Prunet.2022} correspond to an unconverged version of the neutral translation mode. 
We demonstrate that higher numerical fidelity reduces the drift of the neutral mode away from the origin of the complex frequency plane. 
Such explorations are straightforward with \linearresponse\@.

In the case of the Plummer model, we perform our own comparison with $N$-body simulations performed using \EXP\@.
For the radial orbit instability, we find quantitative agreement between predictions and measurements in simulations. 
This agreement opens avenues for validating $N$-body simulations with linear response theory, as well as for studying instabilities in systems where linear response may be complicated. 
In the case of an isotropic \DF\@, we find the same unconverged neutral mode as in the isochrone model. 
We find no evidence for other modes. 
Finally, the $N$-body simulations of the isotropic model suggest that neutral translation modes may not be easily excited in a cluster's lifetime.

The extension of \citetalias{Fouvry.Prunet.2022}'s work to the stability analysis of razor-thin discs is convincingly validated against known semi-analytical results in Mestel discs~\citep{Zang.1976,Evans.Read.1998.II}.
In particular, we satisfyingly retrieve both the frequency and shape of the ${\ell\!=\!2}$ growing mode from~\cite{Zang.1976}. 
Accurately computing the linear response of these dynamically cold systems is key to appropriately studying their long-term relaxation.

All these promising results pave the way for a more systematic and rigorous exploration of linear response for a wider class of clusters and discs, including flattened or indeed triaxial systems. 
For example, radial orbit instabilities have long been suspected as being important for constraining dynamical models of elliptical galaxies. 
With novel data products such as integral field unit spectroscopy \citep[e.g., the MaNGA survey described in][]{Bundy.MaNGA.2015}, new constraints may be available for the orbital content of elliptical galaxies, using modal analysis with \linearresponse\@.

The calculations performed in this paper may straightforwardly be extended to various other cored models and \DFs\@: this is one of the main gains from this generic toolbox. 
Similarly, the software may also be used for other generic computations, such as computing orbital elements in various potentials. 
We hope that these libraries will serve as a base for further calculations and extensions, with the means to approach the speed of fully compiled codes and the readability of scripting languages.

Ultimately, one could eventually build the following other extensions of the software:
(i) extending the mapping to cuspy profiles for which the range of frequencies is unbounded;
(ii) generalising the mapping to integrable rotating spherical clusters or flattened ones;
(iii) accounting for the contributions from branch cuts;
(iv) implementing analytic continuation via rational functions, following \cite{Weinberg.1994}~\citepalias[see appendix~{F} in][for a review]{Fouvry.Prunet.2022};
(v) building the self-gravitating amplification kernel in the time domain rather than the frequency domain~\citep[see, e.g.,][and references therein]{Rozier.etal.2022,Dootson.Magorrian.2022}.

As for long-term relaxation, \linearresponse\ may also be used on various fronts, for example to estimate torques that external perturbations apply on stellar systems.
Indeed, the celebrated LBK formula~\citep{LBK.1972} integrates over resonance conditions.
Phrased differently, it only requires one to account for the resonance part of Landau's prescription for neutral frequencies: the calculation precisely implemented here.
Finally, \linearresponse\ can also easily be adapted to compute the secular response of discs and spheres, through the drift and diffusion coefficients the Balescu--Lenard equation~\citep[see, e.g.,][]{Hamilton.etal.2018}, or to explore their adiabatic response as one slowly varies the mean field or the stellar cluster's \DF\@~\citep[see, e.g.,][]{Reddish.2022}.

%%%%%%%%%%%%%%%%%%%%
\section*{Data Availability}
%%%%%%%%%%%%%%%%%%%%

The \julia\ libraries developed for this paper, \linearresponse, \orbitalelements, \astrobasis\ and \fht\@, are freely available on GitHub.
The \EXP\ simulations used for comparison are available upon reasonable request to the corresponding author.

%%%%%%%%%%%%%%%%%%%%
\section*{Acknowledgements}
%%%%%%%%%%%%%%%%%%%%
We thank  S.~Rozier, M.~Weinberg, C.~Hamilton, S.~de Rijcke, J.~Pe\~{n}arrubia, and A.~Naik for insightful conversations.
We also thank the Kavli Institute for Theoretical Physics for hosting the workshop ``Connecting Galaxies to Cosmology at High and Low Redshift'' and the Higgs Centre for Theoretical Physics for hosting the workshop ``Secular evolution of self-gravitating systems''.
This work is partially supported by the French Agence Nationale de la Recherche under Grant \href{https://www.secular-evolution.org}{SEGAL}  ANR-19-CE31-0017 and by the National Science Foundation under Grant No. NSF PHY-1748958.
We thank Stéphane Rouberol for the smooth running of the Infinity cluster, where the majority of the simulations were performed.
This project is primarily built in the \julia\ language~\citep{julia.language}. 
We used \texttt{scipy} \citep{Scipy.2020} to fit the wave kinetic equation solution.
For visualisations, we used \texttt{iPython} \citep{iPython}, \texttt{numpy} \citep{numpy}, and \texttt{matplotlib} \citep{matplotlib}.

\bibliography{main}

\appendix

\input{Appendices}

\label{lastpage}
\end{document}

%% file: tables/Table1.tex
\begin{table}
\centering
  \begin{tabular}{l|l|l|}
  $\rra / \brc$ & $N$-body $\langle\gamma\rangle / \Omega_{0}$  & Linear response $\langle\gamma\rangle / \Omega_{0} $ \\
    \hline
    \hline
0.75 & 0.046$\pm$0.002 & 0.048$\pm$0.001\\
0.80 & 0.038$\pm$0.002 & 0.038$\pm$0.001\\
0.85 & 0.033$\pm$0.003 & 0.029$\pm$0.001\\
0.90 & 0.023$\pm$0.003 & 0.020$\pm$0.001\\
\end{tabular}
  \caption{Comparison between the measured and predicted growth rates for the radial orbit instability in the Plummer sphere for various normalised anisotropy radii, $\rra/\brc$. 
  The scale frequency, $\Omega_{0}$, is defined for the Plummer cluster (equation~\ref{eq:plummeromega0}).
  Uncertainties on the $N$-body runs arise from variations among different realisations (i.e., initial conditions). 
  Predictions are for the Fiducial configuration, with uncertainties measured from different configurations (Table~\ref{tab:ROIResults}).
  Measurements and predictions satisfyingly agree and retrieve that the smaller the anisotropy radius, the stronger the instability.
  }
  \label{tab:Table1}
\end{table}

%% file: tables/Table2.tex
\begin{table}
\centering
  \begin{tabular}{l|l|l}
    Cut-in index $N$ & This work & \cite{Evans.Read.1998.II} \\
    \hline
    \hline
4   &  $0.878 + 0.126\ri$   & $0.879 + 0.127\ri$    \\
6   & $0.901 + 0.221\ri$    & $0.902 + 0.222\ri$   \\
8   & $0.922 + 0.265\ri$    & $0.922 + 0.266\ri$    \\
\end{tabular}
  \caption{Complex frequency, $\omegaM / \Omega_0$, of the most unstable ${ \ell \!=\! 2 }$ mode
  of truncated Zang discs (${ q \!=\! 6 }$),
  as one varies the index of the inner taper, $N$,
  with $\Omega_0$ the disc's scale frequency (see Appendix~\ref{app:DiscsDF}).
  Here, we compare the numerical prediction from \linearresponse\ 
  with the values from \protect\cite{Evans.Read.1998.II}.
  We refer to Appendix~\ref{app:DiscsDF} for detailed parameters.
  Both predictions satisfactorily agree within $1\%$ precision. The results prove robust to variations of the control parameters (Table~\ref{tab:ZangVariance} for the ${N\!=\!4}$ disc).
}
  \label{tab:ZangModes}
\end{table}

%% file: Appendices.tex
%%%%%%%%%%%%%%%%%%%%
\section{Linear Response Software}
\label{app:Software}
%%%%%%%%%%%%%%%%%%%%

%%%%%%%%%%%%%%%%%%%%
\subsection[OrbitalElements.jl: Orbit Frequency Computation]{\orbitalelements\@: Orbit Frequency Computation}
\label{app:ComputingFrequencies}
%%%%%%%%%%%%%%%%%%%%

% 
% best solution so far: https://tex.stackexchange.com/questions/539403/how-to-use-typewriter-fonts-in-math-mode

% generated by tests/EdgeValidation.jl
% plotted with tests/PlotValidation.py

The \julia\ library \orbitalelements\ is optimised to calculate high-precision quantities for orbits in a static central potential, ${\psi\!=\!\psi(r)}$. 
For convenience, we do not use more generic libraries \texttt{galpy} \citep{Bovy.2015}, \texttt{gala} \citep{PriceWhelan.2017} or \texttt{AGAMA} \citep{Vasiliev.2019}, but rather design our own specific library tailored for our needs.
At the heart of \orbitalelements\ is the ability, given a central potential and its two first derivatives, to convert nearly seamlessly between different orbital elements, i.e., different constants of motion, namely
\begin{itemize}
    \item the pericentre and apocentre radii $(\rperi,\rapo)$;
    \item the effective semi-major axis and eccentricity
    \begin{equation}
        \label{eq:AE}
        a = \frac{\rperi+\rapo}{2} , \quad
        e = \frac{\rapo-\rperi}{\rapo+\rperi};
    \end{equation}
    \item the energy and angular momentum~\citep[equation~{1.3} in][]{LyndenBell.2015}
    \begin{equation}
        \label{eq:EL}
        E \!=\! \frac{\rapo^2 \psi(\rapo) \!-\! \rperi^2\psi(\rperi)}{\rapo^2 \!-\! \rperi^2} , \,
        L \!=\! \bigg[ \frac{2(\psi(\rapo) \!-\! \psi(\rperi))}{\rperi^{-2} \!-\! \rapo^{-2}} \bigg]^{1/2}\!\!\!  ;
    \end{equation}
    \item the actions $(J_r,L)$, where 
    \begin{equation}
        \label{eq:Jr}
        J_r = \displaystyle\frac{1}{\pi}\int_{\rperi}^{\rapo} \!\! \rd r \, v_r,
    \end{equation}
    with $J_{r}$ the radial action and ${v_r\!=\!\sqrt{2(E\!-\!\psi(r))\!-\!L^2/r^2}}$ the instantaneous radial velocity;
    \item the orbital frequencies $(\Omega_1,\Omega_2)$ associated respectively with the radial and azimuthal oscillations;
    \item the frequency ratios $(\alpha,\beta)$ from which the frequencies are computed \citep[see equations~{28} and {29} in][]{Tremaine.Weinberg.1984}
    \begin{equation}
        \label{eq:AlphaBeta}
            \frac{1}{\alpha} = \frac{\Omega_0}{\Omega_1} = \displaystyle \frac{1}{\pi}\int_{\rperi}^{\rapo}  \frac{\rd r}{ v_r}, \quad
            \beta = \frac{\Omega_2}{\Omega_1} =\displaystyle \frac{L}{\pi}\int_{\rperi}^{\rapo} \frac{\rd r}{r^2 v_r},
    \end{equation}
    with $\Omega_{0}$ some given frequency scale (typically the central radial frequency in cored profiles);
    \item the resonance-specific (i.e., dependent on $\bn$) coordinates $(u,v)$ from \citetalias{Fouvry.Prunet.2022} (appendix~{B} therein).
\end{itemize}
In practice, \orbitalelements\ is centred around the effective semi-major axis and eccentricity $(a,e)$, but straightforward conversions between different orbital labels exist as simple function calls. 
These change of coordinates are a requirement for the linear response of self-gravitating systems. 
Indeed, by construction, it involves scanning the full orbital space and dealing appropriately with resonant denominators, as visible in equation~\eqref{eq:ResponseMatrix}.
In \citetalias{Fouvry.Prunet.2022}, these conversions were performed analytically for the isochrone model. 
With this library, we provide a generic computation of orbital elements for any central potential.\footnote{In practice, the library is currently limited to cored potentials. Its extension to cuspy potentials is left for future work.}

%%%%%%%%%%%%%%%%%%%%
\subsubsection{Forward mappings}
\label{app:ForwardMappings}
%%%%%%%%%%%%%%%%%%%%

The direct calculations compute orbital elements (frequency ratios, radial action, etc.) through integrals of the form ${ \int_{\rperi}^{\rapo} \! \rd r ... }$
However, due to the diverging integrand ${1/v_r}$ in equations~\eqref{eq:AlphaBeta}, the radius $r$ is not an appropriate integration variable. 
To improve numerical accuracy, we cure divergences using a mapping from equation~{51} in~\citet{Henon.1971}. 
It reads
\begin{equation}
    \label{eq:Henon_mapping}
    r(w) = a\left[1+ef(w)\right]; \; f(w) = w\left(\tfrac{3}{2} - \half w^2\right),
\end{equation}
with ${w\!\in\![-1,1]}$ the ``H\'{e}non anomaly'', and ${w\!=\!-1}$ (resp.~${w\!=\!1}$) corresponding to the pericentre (resp.\ apocentre).
Using this variable, the integrand
\begin{equation}
    \Theta(w) = \frac{\rd r / \rd w}{v_r(w)},
    \label{eq:Theta}
\end{equation} 
is no longer divergent and equations~\eqref{eq:AlphaBeta} simply read
\begin{subequations}
    \label{eq:AlphaBeta_w}
    \begin{align}
    \frac{1}{\alpha} &= \frac{1}{\pi}\int_{-1}^1 \rd w \, \Theta(w),
    \label{eq:Int_Omega1}
    \\
    \beta &= \frac{L}{\pi}\int_{-1}^1 \rd w \, \frac{\Theta(w)}{r^2(w)}.
    \label{eq:Int_Omega2}
    \end{align}
\end{subequations}

\begin{figure*}
    \begin{center}
        \includegraphics[width=0.9\textwidth]{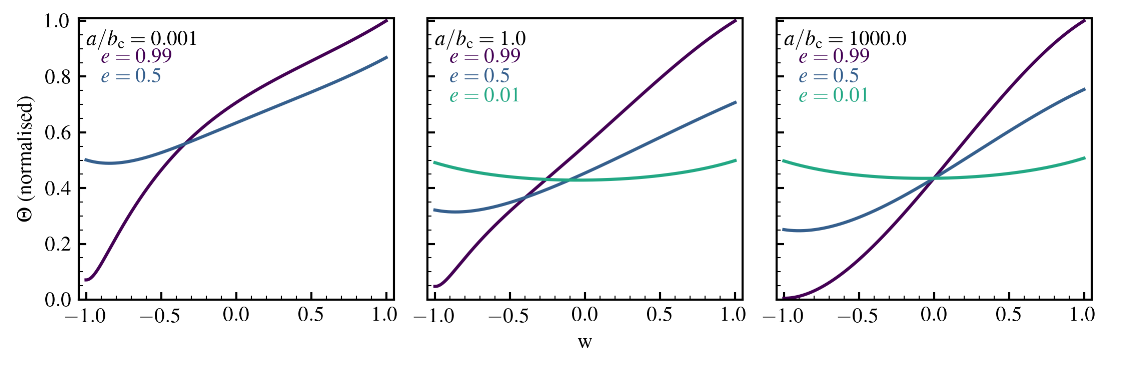}
        \caption{\label{fig:FigureA1} 
        Example of $\Theta(w)$ (equation~\ref{eq:Theta}) for different $(a,e)$ in the isochrone potential, normalised to ${ \Theta (w \!=\! 1) \!=\! 1 }$, and with $\brc$ the potential scale radius.
        The range of $w$ runs from -1 (pericentre) to 1 (apocentre). 
        The curves are smooth and straightforward to integrate via low-order schemes.
        The line for ${ a/\brc \!=\! 0.001 }$ and ${ e \!=\! 0.1 }$ is not shown because it corresponds to an orbit for which interpolation is used (see Appendix~\ref{app:ForwardMappings}).
        }
    \end{center}
\end{figure*}

In practice, the evaluation of $\Theta$ can become numerically unstable close to the boundaries ${w\!=\!\pm 1}$.
To cure this, we perform a second-order expansion for values of $w$ closer than some parameter \texttt{EDGE}. 
Values at the border are obtained from straightforward expansions \citep[see, e.g., equations~53-54 in ][]{Henon.1971}.
The radial action, $J_{r}$, is readily and easily computed by integrating equation~\eqref{eq:Jr} over $w$.
In Fig.~\ref{fig:FigureA1}, we illustrate the function ${ w \!\mapsto\! \Theta (w) }$ for the isochrone potential. 
For most orbits, $\Theta (w)$ is a smooth function that can be integrated using low-order schemes. 
We use the Simpson's 1/3 rule with a parameter \texttt{NINT} to control the number of integration nodes.\footnote{In practice, the smoothness of the function dictates the error in the resulting calculations, rather than the integration scheme.}
%%%%%%%%%%%%%%%%%%%%
%%%%%%%%%%%%%%%%%%%%

For near-circular orbits, i.e., ${e\!\to\!0}$, and/or very small semi-major axes, i.e., ${a\!\to\!0}$, the evaluation of $\Theta (w)$ can break down.
To tackle this issue, we use analytic expressions for circular (${e\!=\!0}$) and central (${a\!=\!0}$) orbits, and then perform a second-order expansion for values close to the borders. 
For the sake of numerical stability, similar expansions are performed close to radial orbits, i.e., ${e\!\to\!1}$. 
In that case, the values on the border are computed through usual integrations (there exist no analytical expressions) before being interpolated.
The same interpolations are used to compute the orbit's energy and actions.
In practice, the regions of expansions are set by the parameters \texttt{TOLECC} and \texttt{TOLA} (rescaled by the characteristic radius parameter \texttt{rc}).

%%%%%%%%%%%%%%%%%%%%
\subsubsection{Backward mappings}
\label{app:BackwardMappings}
%%%%%%%%%%%%%%%%%%%%

We now have at our disposal forward transformations from semimajor axis and eccentricity to energy, actions and frequencies (ratios). 
These mappings are not analytically invertible. 
We therefore employ a Newton--Raphson descent to invert them, e.g., to construct the function ${(\alpha,\beta)\!\mapsto\!(a,e)}$. 
It requires the knowledge of the forward mapping Jacobians, i.e., sets of derivatives. 
These derivatives are estimated via simple two-point finite differences, on the scales \texttt{da} and \texttt{de}. 
Finally, the Newton--Raphson algorithm is controlled with a maximal number of iterations (\texttt{ITERMAX}) and a accuracy goal (\texttt{inv$\varepsilon$}). 
The effective accuracy of the  inversion is mainly determined by the orbital integration (\texttt{NINT}) and the finite differences (\texttt{da} and \texttt{de}).

%%%%%%%%%%%%%%%%%%%%
\subsubsection{Resonances variables}
\label{app:ResonancesVariables}
%%%%%%%%%%%%%%%%%%%%

Let us now discuss the computation of resonance-specific (i.e., dependent on $\bn$) coordinates $(u,v)$. 
Within these coordinates, resonances lines (${\omega_{\bn} \!=\! \bn \!\cdot\!\bOmega\!=\!\mathrm{cst.}}$) become straight lines (${u\!=\!\mathrm{cst.}}$).
Their construction from the frequency ratios, $(\alpha,\beta)$, is concisely presented in appendix~{B} of \citetalias{Fouvry.Prunet.2022}. We refer to this paper for definitions and notation.

In \orbitalelements\@, we extend the resonance-specific mapping to situations where the potential and \DF\ are no fully self-consistent.
In this situation, a significant portion of the frequency space might be unimportant for the system's linear response.
This is for example the case in tapered Zang discs (Appendix~\ref{app:DiscsDF}), where the frequency profile diverges in the ``unpopulated'' center.

We therefore truncate the domain in ${(\alpha,\beta)}$ effectively explored.
This does not affect the computation of the frequencies, $\bOmega$, but only the definition of the resonance variables ${(u,v)}$.
This effective truncation is set by the two parameters \texttt{rmin} and \texttt{rmax}.
Then, the effective domain in ${(\alpha,\beta)}$ is restricted to the region between ${\alphamin\!=\!\alphac(\texttt{rmax})}$ and ${\alphamax\!=\!\alphac(\texttt{rmin})}$ with ${ \alphac (r) \!=\! \alpha (a \!=\! r , e \!=\! 0) }$ the (outward decreasing) circular frequency ratio.
For example, in the case of a Mestel disc (Appendix~\ref{app:DiscsDF}), ${\texttt{rmin}\!>\!0}$ ensures that the frequency support is finite, and that the $(u,v)$ domain is focused on orbital regions effectively populated.

Fortunately, compared to \citetalias{Fouvry.Prunet.2022}, adding these truncations only amounts to slightly modifying the boundary values of the resonance frequency ${\omega_{\bn}\!\in\![\omeganm,\omeganp]}$ and the variable ${v\!\in\![\vnm,\vnp]}$.
More precisely, the extrema of ${\omega_{\bn}}$ \citepalias[see equation~{B7} in][]{Fouvry.Prunet.2022} are now reached either in the four edges ${(\alphamin,\half)}$, ${(\alphamin,\betac[\alphamin])}$, ${(\alphamax,\half)}$, ${(\alphamax,\betac[\alphamax])}$ or along the curve ${(\alpha,\betac [\alpha])}$ with ${\alphamin\!\leq\!\alpha\!\leq\!\alphamax}$.
For the $v$ variable, the first two constraints of equation~{(B10)} in \citetalias{Fouvry.Prunet.2022} become ${\alphamin\!\leq\!v; \, v\!\leq\!\alphamax}$.
Finally, the same procedure as in \citetalias{Fouvry.Prunet.2022} is used to obtain the boundary values.

For a fully self-gravitating system (e.g., a Plummer sphere as in Appendix~\ref{app:Plummer}), one can stick to the default parameters values which do not introduce such domain restriction.

%%%%%%%%%%%%%%%%%%%%
\subsubsection{Parameters}
\label{app:OrbitalParameters}
%%%%%%%%%%%%%%%%%%%%

%%%%%%%%%%%%%%%%%%%%
\input{tables/TableA1}
%%%%%%%%%%%%%%%%%%%%
For the user that wants straightforward simplicity, control parameters are largely hidden, and set to tested defaults. Nonetheless, we also provide a documented interface to change them.
Indeed, all exposed function calls allow the user to modify control parameters, through an optional final argument. 
The control parameters are summarised in Table~\ref{tab:OEparams}, but are described in more detail throughout the previous subsections.

%%%%%%%%%%%%%%%%%%%%
\subsubsection{Tests}
\label{app:OrbitalTests}
%%%%%%%%%%%%%%%%%%%%

\begin{figure}
    \begin{center}
        \includegraphics[width=0.45\textwidth]{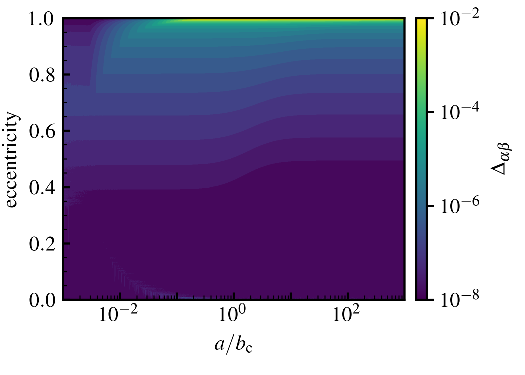}
        \caption{\label{fig:FigureA2} 
        Accuracy of the numerical frequency calculation for the (analytical) isochrone model, using the distance $\Delta_{\alpha\beta}$ (equation~\ref{eq:def_Delta}).
        Extremely radial orbits are the most difficult ones to deal with.
        }
    \end{center}
\end{figure}

\begin{figure}
    \begin{center}
        \includegraphics[width=0.45\textwidth]{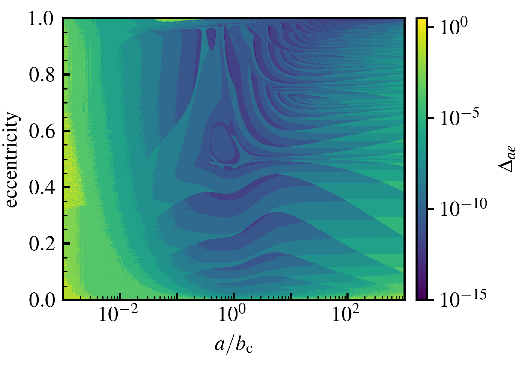}
        \caption{\label{fig:FigureA3} 
        Accuracy of the numerical inversion from ${\left(\alpha,\beta\right)\!\mapsto\!(a,e)}$ for the Plummer model. 
        For each grid point in ${(a,e)}$, we first compute the frequencies ${(\alpha,\beta)}$, then ``invert'' these frequencies back to calculate ${(\tilde{a},\tilde{e})}$. We define a relative accuracy metric ${\Delta_{ae}}$ in equation~(\ref{eq:inversionaccuracyae}), which is the colourmap. 
        The effective inversion is stable and shows a very satisfactory overall performance. 
        The accuracy owes primarily to the recovery of eccentricity; semi-major axis is generally recovered to a higher precision.
        }
    \end{center}
\end{figure}

Figs.~\ref{fig:FigureA2} and \ref{fig:FigureA3} demonstrate the fidelity of our orbit frequency calculation and subsequent inversion. 
We use the default control parameters in \orbitalelements\ for this accuracy benchmark.

First, in Fig.~\ref{fig:FigureA2}, we test the accuracy of $\alpha$ and $\beta$ calculations for the isochrone model (for which we know the analytic values of $\alpha$ and $\beta$). 
As a summary statistics of the accuracy, we define the distance between the numerically-calculated values of $\alpha$ and $\beta$ (denoted $\tilde{\alpha}$, $\tilde{\beta}$) and the analytic values (simply denoted $\alpha$, $\beta$) through
\begin{equation}
    \label{eq:def_Delta}
    \Delta_{\alpha\beta} = \sqrt{\left( |\tilde{\alpha}-\alpha| / \alpha \right)^2 + \left( |\tilde{\beta}-\beta | / \beta \right)^2}.
\end{equation}
In practice, $\Delta_{\alpha\beta}$ is dominated by the accuracy contribution from $\tilde{\beta}$. 
To probe the full range of relevant semi-major axes, we sample $a$ using 1000 log-spaced points ${a/\brc\!\in\![10^{-3},10^3]}$, and 1000 linear-spaced points ${e\!\in\![0,1]}$.
The accuracy is better than 1\% at all points sampled, and in most cases is better than 0.001\% accurate.
%%%%%%%%%%%%%%%%%%%%
%%%%%%%%%%%%%%%%%%%%

In Fig.~\ref{fig:FigureA3}, we use the same ${1000\!\times\!1000}$ grid of $(a,e)$ to test the inversion from frequencies back to $(a,e)$. 
This time we use the Plummer potential, to test both the numerical computation of ${(a,e)\!\mapsto\!(\tilde{\alpha},\tilde{\beta})}$ and the subsequent numerical inversion ${(\tilde{\alpha},\tilde{\beta})\!\mapsto\!(\tilde{a},\tilde{e})}$. 
The figure shows a combination of the relative difference ${|\tilde{a}\!-\!a|/a}$ and absolute difference ${|\tilde{e}\!-\!e|}$ for each point in the grid, given by
\begin{equation}
    \Delta_{ae} = \sqrt{\left(|\tilde{a}-a| / a \right)^2 + \left(|\tilde{e}-e|\right)^2}.
    \label{eq:inversionaccuracyae}
\end{equation}
In practice we find that the inversions are quite accurate, with typical differences on the order of 0.001\%. 
However, in certain regions it becomes difficult to accurately recover the input $(a,e)$ values, namely orbits close to radial or circular, and orbits with small semi-major axes. 
Fortunately, these inaccuracies have little impact on later calculations. 
%%%%%%%%%%%%%%%%%%%%
%%%%%%%%%%%%%%%%%%%%

In \orbitalelements\@, we have taken care to optimise the calculation time and memory allocations. 
While individual frequency calculations are completed in \texttt{NINT} integration steps, the inversion to $(a,e)$ can take up to \texttt{ITERMAX} longer. 
As a testimony of performance, the grid of ${10^6}$ ${(a,e)}$ values and their inversions in Fig.~\ref{fig:FigureA3} took ${\sim\!140\mathrm{s}}$ on a single core of an M1 Macbook Air.

%%%%%%%%%%%%%%%%%%%%
\subsection[AstroBasis.jl: Basis Computation]{\astrobasis\@: Basis Computation}
\label{app:ComputingBasis}
%%%%%%%%%%%%%%%%%%%%

Fundamental to the matrix method are the chosen basis functions.
The \astrobasis\ library is an implementation of several different radial basis functions, ${r\!\mapsto\!U^{\ell}_p(r)}$, with a straightforward interface.\footnote{Generically, the radial functions, $U_{p}^{\ell} (r)$, can always be taken to be real~\citep[see section~{3.2} in][]{Heyvaerts.2010}. Conveniently, this makes ${ G(u) }$ a purely real function, as in equation~\eqref{eq:FHT}.}
At present, \astrobasis\ supports the bases from \citet{CluttonBrock.1972}, \citet{CluttonBrock.1973}, \citet{Kalnajs.1976}, \citet{Fridman.Polyachenko.1984}/\citet{Weinberg.1989} (Bessel), and \citet{Hernquist.Ostriker.1992}.
The spherical calculations in this paper use the basis from \citet{CluttonBrock.1973}, which is re-described in appendix~{B} of \citet{Fouvry.etal.2021}. The associated lowest-order function matches the Plummer potential exactly.
The disc calculations in this paper use the basis from~\citet{CluttonBrock.1972}.

The user must choose parameters to define the basis, including the gravitational constant $G$ (\texttt{G}), the scaling radius for the basis $\rrb$ (\texttt{rb}), the maximum harmonic number ${\ell_{\mathrm{max}}}$ (\texttt{lmax}), and the maximum radial number $n_{\rm max}$ (\texttt{nradial}).
Basic parameters are listed in Table~\ref{tab:ABparams}. Some bases may require additional parameters.
%%%%%%%%%%%%%%%%%%%%
\input{tables/TableA2}
%%%%%%%%%%%%%%%%%%%%

When defining a basis, the prefactors are computed in advance and tabulated so that calls to evaluate the basis are rapid. The functions at a given radius are evaluated on the fly. 
For bases computed by recursion, all radial orders are computed in one step to optimise memory usage. 

\begin{figure}
    \begin{center}
        \includegraphics[width=0.50\textwidth]{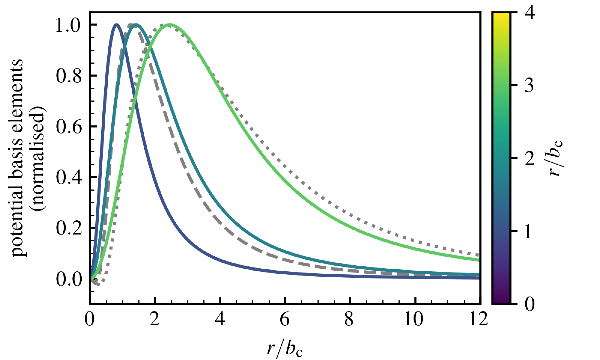}
        \caption{\label{fig:FigureA4} 
        Lowest-order ${\ell\!=\!2}$ radial basis function for the \citet{CluttonBrock.1973} basis as a function of radius.
        Three different values of \texttt{rb} are shown, ${\rrb/\brc\!=\![1,\sqrt{3},3]}$.
        The dashed grey curve is the ROI mode for the ${\rra/\brc\!=\!0.75}$ Plummer model (see Fig.~\ref{fig:Figure1}), which peaks at the same location as the ${\rrb/\brc\!=\!\sqrt{3}}$ basis.
        The dotted grey curve is the ROI mode for the ${\rra/\brc\!=\!1.0}$ isochrone model, which peaks at the same location as the ${\rrb/\brc\!=\!3}$ basis.
        Using a basis tailored for the underlying mean potential, as is the case here, greatly improves the convergence of the linear response prediction.}
    \end{center}
\end{figure}

Figure~\ref{fig:FigureA4} shows an example of basis functions from \astrobasis\@. 
In this case, we compare the shape of the  first $\ell\!=\!2$ element of \citet{CluttonBrock.1973}'s  basis to the shape of the ROI modes from the isochrone cluster (Appendix~\ref{app:IsochroneROI}) and the Plummer cluster (Section~\ref{sec:PlummerROI}).
To give a sense of the similarity of the basis function to the predicted ROI mode shape, we chose the \texttt{rb} value to match the peak of the predicted shape for the mode.
In general, for linear response calculations, one decides the maximum order for the expansion to resolve the expected structure of the mode.
In many cases, the user will not have an a priori guess for the shape of the mode.
This entails experimentation with \texttt{rb} and \texttt{nradial}.
%%%%%%%%%%%%%%%%%%%%
%%%%%%%%%%%%%%%%%%%%

In addition to analytic bases for a select few models, recent basis function implementations have created an opportunity for tailored basis functions \citep{Petersen.Weinberg.Katz.2022,Lilley.vandeVen.2023}.
The applicability of these tailored basis functions to matrix method calculations should be explored further in future work.

%%%%%%%%%%%%%%%%%%%%
\subsection[FiniteHilbertTransform.jl: Specialised Legendre Integration]{\fht\@: Specialised Legendre Integration} 
\label{app:LegendreIntegration}
%%%%%%%%%%%%%%%%%%%%

\citetalias{Fouvry.Prunet.2022} describes a method to adapt Landau's integral prescription to the case of self-gravitating systems in order to approximate the integral\footnote{Equation~\eqref{eq:FHT} is the finite-interval version of the Hilbert transformation~\citep{Tricomi.1951}, hence the name of this software.}
\begin{equation}
    M (\omega) = \intLb \rd u \frac{G (u)}{u - \omega},
    \label{eq:FHT} 
\end{equation} 
at all points in the complex plane, ${ \omega \!\in\! \mathbb{C} }$,
using Legendre polynomials, $P_{k}(u)$. 
We note that we must compute integrals of this form for each resonance $\bn$ and basis element pairs ${ (p,q) }$, as highlighted in equation~\eqref{eq:M_int_u}. 

To compute equation~\eqref{eq:FHT}, we first approximate its numerator via
\begin{equation}
G(u) \!\simeq\! \sum_{k = 0}^{K_{u}-1} a_{k} \, P_{k} (u) ,
\label{eq:approx_Gu}
\end{equation}
where, importantly, $K_u$ (defined as \texttt{Ku} in \fht\@) controls the quality of the approximation.
In practice, the coefficients $a_{k}$ entering equation~\eqref{eq:approx_Gu} are estimated through a Gauss--Legendre quadrature with $K_{u}$ nodes (see equation~{D3} in~\citetalias{Fouvry.Prunet.2022}).\footnote{For smooth enough $G(u)$, this is a robust approximation scheme~\citep{Trefethen2019}.}

Then, the integral from equation~\eqref{eq:FHT} is computed through
\begin{equation}
    M (\omega) = \sum_{k = 0}^{K_u - 1} a_{k}  D_{k} (\omega) ,
    \label{eq:Msum}
\end{equation} 
where 
\begin{equation}
    D_{k} (\omega) =  \intLb  \rd u  \frac{P_{k} (u)}{u - \omega}.
    \label{def_D}
\end{equation}
The function $D_k(\omega)$ explicitly implements Landau's prescription and is defined in appendix~{D} of~\citetalias{Fouvry.Prunet.2022}. 
In brief, $D_k(\omega)$ is proportional to Legendre functions of the second kind when $\ImPart [\omega]>0$, with an added contribution when $\ImPart[\omega]\le0$ \citep[cf.\ section~{5.2.4} of][]{Binney.Tremaine.2008}. 
The library \fht\ implements the computation of (i) the $a_k$ coefficients for a given function $G$, and (ii) the ${\omega\!\mapsto\!D_k(\omega)}$ functions from equation~\eqref{eq:Msum} at all points in the complex plane.

The present method has mainly been introduced to probe linear response for damped frequencies, i.e., in the lower-half of the complex plane. 
In this regime, Landau's prescription requires to give a meaning to $G(\omega)$ with ${\omega\!\in\!\mathbb{C}}$.
Phrased differently, one has to perform an analytical continuation of these $G$ functions, which involve intricate, non-analytically known functions in the present self-gravitating case.
Given that analytical continuation is intrinsically a (severely) ill-conditioned numerical problem \citep{Trefethen.2020}, for damped frequencies, i.e., negative $\ImPart[\omega]$, the effective numerical precision plays an important role in setting the floor for accuracy.

\begin{figure}
    \begin{center}
        \includegraphics[width=0.50\textwidth]{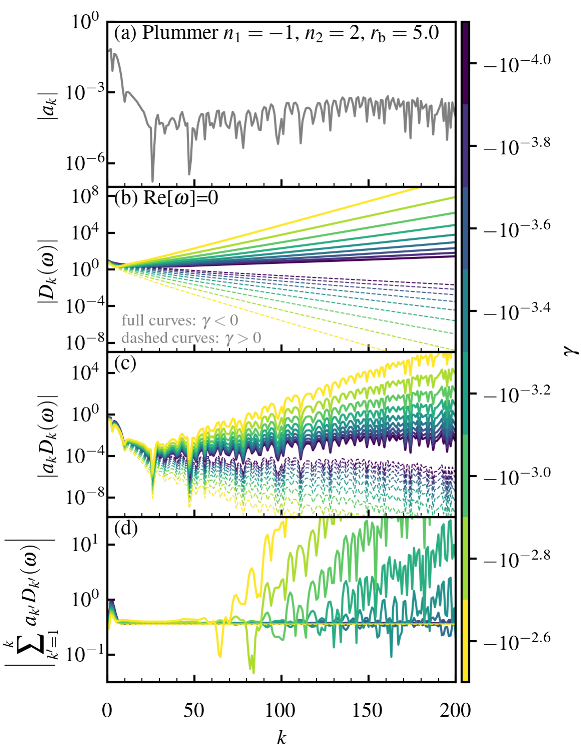}
        \caption{\label{fig:FigureA5} 
        Convergence of finite Hilbert Transform integral calculation for the ${\ell\!=\!2}$ radially-anisotropic Plummer cluster from Fig.~\ref{fig:Figure1},
        fixing ${ \rra / \brc \!=\!1.0}$, ${ \bn \!=\! (-1,2) }$,
        ${ p \!=\! q \!=\! 1 }$,
        and using the Fiducial configuration settings.
        Here, the response matrix is evaluated for ${ \omega\!=\!0\!+\! \ri \gamma }$, with ${\gamma<0}$ (i.e., damped frequencies; solid curves) and zero oscillation frequency.
        Dashed curves show the corresponding calculation for ${-\gamma}$, i.e., unstable frequencies.
        Panel (a) shows that the coefficients $a_k$. They decay for small $k$, but they eventually saturate, as a result of finite numerical accuracy.
        Panel (b) shows that for all ${ k \!>\! 20 }$, the quantity $D_k$ increases exponentially as a function of $k$.
        Panel (c) shows the individual components of $a_kD_k$ as a function of $k$.
        For $k$ large enough,
        this product does not decay anymore.
        Finally, panel (d) shows the cumulative sum in equation~\eqref{eq:Msum} as a function of $k$.
        For $\gamma$ sufficiently close to the real frequency line, the sum is convergent.
        However, as $\ImPart[\omega]$ becomes more negative, the sum begins to diverge for smaller and smaller $k$.
        In contrast to the solid lines which show the damped frequencies (i.e.\@, ${ \gamma \!<\! 0 }$), the dashed lines -- which show the unstable frequencies (i.e.\@, ${ \gamma \!>\! 0 }$) -- are always convergent.
        }
    \end{center}
\end{figure}

Indeed, the complex Legendre polynomials, $P_{k} (\omega)$, diverge in the lower-half of the complex frequency plane. 
In practice, this divergence only gets approximately cancelled out by the decaying coefficients $ k \!\mapsto\! |a_{k}| $. 
Due to the limitations of finite numerical precision, this results in ringing and numerical saturation in the lower half of the complex frequency plane (see, e.g., Fig.~\ref{fig:Figure5}).
Figure~\ref{fig:FigureA5} demonstrates why this numerical saturation takes place.
For values of $\ImPart[\omega]$ becoming increasingly more negative, the sum in equation~\eqref{eq:Msum} diverges more and more (panel d), owing to the combination of ultimately non-decreasing $|a_k|$ (panel a), and exponentially increasing ${ |D_k (\omega)| }$ (panel b).
We emphasize that numerical saturation is eventually unavoidable given that the $G$ functions are not analytically known, whatever the chosen method.
%%%%%%%%%%%%%%%%%%%%
%%%%%%%%%%%%%%%%%%%%

Implementation of strategies to mitigate this numerical divergence, i.e., to push the sum from equation~\eqref{eq:Msum} further down in the complex frequency plane, are currently underway.
To this extent, delaying the saturation of the decomposition coefficients, $a_k$, would prove useful. 
It requires an enhanced accuracy on the $u\!\mapsto\!G(u)$ functions, i.e., on (i) the \orbitalelements\ mappings and associated gradients, and (ii) the Fourier transform of basis elements. 
This would expectantly increase precision in both the upper-half and lower-half planes.
Though, these series of coefficients will inevitably saturate at machine precision (at best). 
Therefore, appropriate, a posteriori, regularisation procedures should also be designed to push significantly further down in the complex plane,
e.g.\@, by reducing the effective range of summation over $k$ in equation~\eqref{eq:Msum}.

%%%%%%%%%%%%%%%%%%%%
\subsection[LinearResponse.jl: Linear Response Computation]{\linearresponse\@: Linear Response Computation}
\label{app:ComputingLR}
%%%%%%%%%%%%%%%%%%%%

The computation of the response matrix and associated by-products is performed by \linearresponse\@. 
It mainly requires the user to provide (i) the considered gravitational potential (and two derivatives), (ii) the \DF\ (through its directional derivatives ${\bn\!\cdot\!\p F/\p\bJ}$), and (iii) a bi-orthogonal basis.
Some of these are available via \orbitalelements\ and \astrobasis\@, but the user can easily supply their own.

For a given harmonic $\ell$ and for each resonance $\bn$ and each matrix element $(p,q)$, the calculations proceed in three phases.
The first two aim at computing the ${u\!\mapsto\!G(u)}$ functions involved in equation~\eqref{eq:M_int_u}, namely (i) by computing the Fourier transform of basis elements, and (ii) by performing an integral along the resonance line, i.e., over the resonance variable $v$.
The third phase is to decompose these functions over Legendre polynomials, through the computation of the $a_k$ coefficients from equation~\eqref{eq:approx_Gu} using \fht\@.
Once these coefficients are known, the response matrix can be efficiently computed for any given complex frequency $\omega$.

%%%%%%%%%%%%%%%%%%%%
\input{tables/TableA3}
%%%%%%%%%%%%%%%%%%%%

%%%%%%%%%%%%%%%%%%%%
\subsubsection{Computing the G functions}
\label{app:ComputingG}
%%%%%%%%%%%%%%%%%%%%

Equation~\eqref{eq:ResponseMatrix} reduces to equation~\eqref{eq:M_int_u} when taking 
\begin{equation}
    \label{eq:G_u_J}
    G(u) = \int_{0}^{1} \rd v' \frac{2}{\Omega_0 (\omeganp - \omeganm)}\left| \frac{\p \bJ}{\p (u,v')} \right| G(\bJ[u,v']),
\end{equation}
where $\omeganm$ and $\omeganp$ are introduced in Appendix~\ref{app:ResonancesVariables} and the ${\bJ\!\mapsto\!G(\bJ)}$ functions can be found in Appendix~\ref{app:LRIntegrands}.
In both cases, they involve the Fourier transform of basis elements \citep{Tremaine.Weinberg.1984}.
They read
\begin{equation}
    \label{eq:Wln}
    W^{\ell\bn}_p(\bJ) = \int_{-1}^{1} \rd w \, \frac{\rd \theta_1}{\rd w}\,
    U^{\ell}_p (r) \,
    \cos\!\big(n_1 \theta_1 \!+\! n_2[\theta_2\!-\!\phi]\big),
\end{equation}
where the radius $r$ and the angles $\theta_{1}$ and ${\theta_2\!-\!\phi}$ are implicit functions of the orbit, $\bJ$, and the Hénon anomaly, $w$, introduced in equation~\eqref{eq:Henon_mapping}.
Following appendix~{B} of \cite{Rozier.etal.2019}, we perform the nested integration over $w$ in equation~\eqref{eq:Wln} simultaneously, pushing the angles with 
\begin{equation}
    \frac{\rd}{\rd w} (\theta_1, \theta_2\!-\!\phi) = \left(\Omega_1,\Omega_2\!-\! L / r^{2} \right)\Theta(w) ,
    \label{grad_theta}
\end{equation}
where $\Theta$ follows from equation~\eqref{eq:Theta}.
In practice, we use the RK4 scheme \citep[see, e.g.,][]{Press.2007} with \texttt{Kw} steps.
Importantly, we perform the integration backward from apocentre ${(w,\theta_1,\theta_2\!-\!\phi)\!=\!(1,\pi,0)}$ to pericentre to mitigate errors.

For a fixed accuracy goal, more eccentric orbits ($e\!\rightarrow\!0$) typically require more integration steps.
We therefore add the ability to choose the number of steps adaptively, following
\begin{equation}
    \label{eq:ADAPTIVEKW}
    \texttt{Kw} \leftarrow \lceil \texttt{Kw}/(0.1+(1-e))\rceil.
\end{equation}
This option is set by the boolean parameter \texttt{ADAPTIVEKW}.

Computing the Fourier transforms at each point $(u,v')$ and for each resonance $\bn$ is the most intensive part of the linear response computations.
For the Fiducial configuration settings, each resonance takes approximately 16 seconds on an M1 Macbook Air.
The calculation of the resonances can be parallelised.
Helpfully, these quantities do not depend on the \DF\@ but only on the potential.
They are therefore computed once and stored in HDF5 files for future use.

The integral along the resonance line in equation~\eqref{eq:G_u_J} is performed on a rescaled variable $v'$, compared to the choice ${v\!=\!\alpha}$ (for ${n_2\!\neq\!0}$) used in \citetalias{Fouvry.Prunet.2022}.
Indeed, in practice, we use
\begin{equation} 
    v'=\left[\frac{v-\vnm}{\vnp-\vnm}\right]^\texttt{VMAPN},
    \label{eq:VMAPN}
\end{equation}
with ${v\!\in\![\vnm,\vnp]}$ and \texttt{VMAPN} some integer parameter which allows for the variable to spread more evenly along the resonance line.
This proves particularly useful for the \ILR\@.
Ultimately, the integral over $v'$ is performed using a simple midpoint rule with \texttt{Kv} points.

In practice, each function ${u\!\mapsto\!G^{\ell\bn}_{pq}(u)}$ is evaluated in \texttt{Ku} points, ${\{u_k\}_{k}}$, chosen according to the Gauss--Legendre quadrature.
For the Fourier transform of basis elements, for each resonance $\bn$, we end up with ${\texttt{Ku}\!\times\!\texttt{Kv}}$ vectors $\{W^{\ell\bn}_p(u,v)\}_p$ of length \texttt{nradial}.
Computing these vectors is the most memory- and time-consuming operation.
We therefore cache files for each considered resonance $\bn$, in arrays of size ${\texttt{Ku}\!\times\!\texttt{Kv}\!\times\!\texttt{nradial}}$.
For the $G$ functions, for each resonance $\bn$, we end up with ${\texttt{nradial}\!\times\!\texttt{nradial}}$ vectors ${\{{G}^{\ell \bn}_{pq}(u_k)}\}_k$ of length \texttt{Ku}. 
These are also cached.

%%%%%%%%%%%%%%%%%%%%
\subsubsection{Computing the response matrix}
\label{app:ComputingM}
%%%%%%%%%%%%%%%%%%%%

Once the $G$ functions tabulated, their Legendre decomposition coefficients from equation~\eqref{eq:approx_Gu} are computed using \fht\@.
For each resonance $\bn$, this results in \texttt{Ku} matrices of size ${\texttt{nradial}\!\times\!\texttt{nradial}}$.\footnote{We use the symmetry ${A_{pq}\!=\!A_{qp}}$ to reduce the computational cost.}

Then, for a given complex frequency $\omega$, the response matrix from equation~\eqref{eq:M_int_u} is computed using equation~(\ref{eq:Msum}).
We write
\begin{equation}
    \boldmatrix{M}^{\ell}(\omega) = \sum_{\bn} \sum_{k} D_k(\varpi_{\bn})\,\boldmatrix{A}^{\ell\bn}_k ,
\end{equation}
where ${\omega\!\mapsto\!\varpi_{\bn}(\omega)}$ is given by equation~(11) in \citetalias{Fouvry.Prunet.2022}, and the $D_k$ functions are computed by via \fht\@.

On top of this computation of the response matrix, \linearresponse\ also supports a range of analysis steps, namely:
\begin{enumerate}
    \item Mapping the complex plane by computing the determinant of the (gravitational) dielectric matrix from equation~\eqref{eq:def_dielectric} at a list of $\omega$ values, as illustrated in Fig.~\ref{fig:Figure5}.
    When this determinant vanishes, the system supports a mode.
    \item Identifying the frequency of a mode, $\omegaM$, using a gradient descent algorithm.\footnote{This involves two new control parameters, \texttt{POLETOL} and \texttt{NRSTEP}, and requires the user to define a starting point. In practice, the pole finding only weakly depends on these control parameters, and reasonable defaults have been set.}
    This is the method used in the main text to report on the modes' frequencies.
    \item Identifying the shape of a mode, $\psiM$, given its frequency. 
    We write
    \begin{equation}
        \label{eq:def_Xp}
        \psi_{\mathrm{M}}(\br) = \RePart \bigg[ \sum_{p} X_p \, \psi_p(\br) \bigg] ,
    \end{equation}
    where ${ \{ X_p \}_{p} }$ is the kernel eigenvector of the dielectric matrix, and ${ \psi_p (\br)\!\propto\!U^{\ell}_{p} (r) }$ (with an angular dependence set by the considered geometry) runs over the potential basis functions used in the response matrix computation.
\end{enumerate}

%%%%%%%%%%%%%%%%%%%%
\subsubsection{Linear response parameters}
\label{app:LRparams}
%%%%%%%%%%%%%%%%%%%%

As for \orbitalelements\@, control parameters for \linearresponse\ are set in a \julia\ structure taken as an optional argument of all exposed functions.
In addition, all control parameters from \orbitalelements\ are passed to \linearresponse\ as a structure, owing to the computations offloaded to \orbitalelements\@. 
In general we find that the default settings result in high enough accuracy for our problems of interest. 
The relevant control parameters are summarised in Table~\ref{tab:LRparams}.

The accuracy of \linearresponse\@ is not straightforward to predict.
Let us nonetheless report on some heuristics.
First, increasing the number $\texttt{nradial}$ of basis elements leads to more oscillating radial functions, hence more difficult reconstructions.
This requires therefore a more precise quadrature. 
The same with $\texttt{n1max}$, the maximum radial resonance number.
As it increases, the resonance structures become increasingly complicated and require additional care when estimating integrals.

To conclude this section, let us comment on performance.
The generic computation of the self-gravitating linear response involves numerous  nested integrals, from frequency computations to finite Hilbert transforms.
For these intense computations, \julia\ provides an ideal language to optimise the running time while keeping a high readability. 
As an example, for a given ${(\ell,\bn)}$, a typical computation of all the needed values of ${\{W^{\ell\bn}_{p}(u,v)\}_{p}}$, with ``Fiducial'' parameters and ${\texttt{nradial}\!=\!100}$, takes ${\sim\!40}$ seconds on a single core of an M1 Macbook Air.
Fortunately, parallelising over resonances is straightforward in \julia\@.
Several resonances can be then computed at once, depending on the thread count of the hardware.

%%%%%%%%%%%%%%%%%%%%
\section{Linear response integrands}
\label{app:LRIntegrands}
%%%%%%%%%%%%%%%%%%%%

%%%%%%%%%%%%%%%%%%%%
\subsection{For spheres}
\label{app:G_spheres}
%%%%%%%%%%%%%%%%%%%%

As in~\citetalias{Fouvry.Prunet.2022}, in ${3D}$ the full expression of $\boldmatrix{G}^{\ell\bn}(\bJ)$ in equation~\eqref{eq:ResponseMatrix} is given by 
\begin{equation}
    \label{eq:def_Gpq}
    G^{\ell \bn}_{pq} (\bJ) \!=\! - \frac{2 (2 \pi)^{3}}{2 \ell + 1} \big| y_{\ell}^{n_{2}} \big|^{2}  L\, \left(\bn \!\cdot\! \frac{\p F}{\p \bJ}\right) W_{p}^{\ell\bn} (\bJ)  W_{q}^{\ell\bn} (\bJ) ,
\end{equation}
where we introduced ${y_{\ell}^{n}\!=\!Y_{\ell}^{n}(\tfrac{\pi}{2},0)}$, with ${Y_{\ell}^{m}(\vartheta,\phi)}$ the spherical harmonics normalised so that ${\!\int\! \rd\vartheta \rd\phi \sin\!(\vartheta) |Y_{\ell}^{m}(\vartheta,\phi)|^{2} \!=\! 1}$. 
Further details may be found in~\citetalias{Fouvry.Prunet.2022}, including the Jacobian for the conversion to $\boldmatrix{G}^{\ell \bn}(u,v)$, needed in practice by \linearresponse\@.
In equation~\eqref{eq:def_Gpq}, the Fourier-transformed basis elements follow from equation~\eqref{eq:Wln}.

%%%%%%%%%%%%%%%%%%%%
\subsection{For discs}
\label{app:G_discs}
%%%%%%%%%%%%%%%%%%%%

It is straightforward to extend the linear response computations of~\citetalias{Fouvry.Prunet.2022} to the case of razor-thin discs with axial symmetry. In this case, the definition of the response matrix~\citep[see section~{5.3} in][]{Binney.Tremaine.2008} immediately applies, without any integral over a third dummy action as in the ${3D}$ spherical case \citep[see section~{4} in][]{Hamilton.etal.2018}.
Then, simply using 
\begin{equation}
    \label{eq:def_Gpq_discs}
    G_{pq}^{\ell\bn}(\bJ) = 
    - (2 \pi)^2 \,\delta_{\ell}^{n_2}\, \left( \bn \!\cdot\! \frac{\p F}{\p\bJ} \right)
    W_{p}^{\ell\bn} (\bJ) \,W_{q}^{\ell\bn} (\bJ),
\end{equation}
instead of equation~\eqref{eq:def_Gpq}, one can compute the linear response of razor-thin discs using all the numerical tools from \linearresponse\@.
Equation~\eqref{eq:def_Gpq_discs} imposes the constraint ${n_2\!=\!\ell}$ and the resonance ${\bn\!=\!(0,0)}$ does not contribute (to the ${\ell\!=\!0}$ response matrix) as in the spherical case.
It also involves the Fourier transform of ${2D}$ bi-orthogonal basis elements, following equation~\eqref{eq:Wln}, which are described in Appendix~\ref{app:2Dbasis}.

%%%%%%%%%%%%%%%%%%%%
\section{Validation with the isochrone sphere}
\label{app:Isochrone}
%%%%%%%%%%%%%%%%%%%%

In this section, we revisit the calculations from~\citetalias{Fouvry.Prunet.2022} for the spherical isochrone potential \citep{Henon.1959}. 
Because the isochrone has analytic expressions for the frequencies, one can readily evaluate integrals to high numerical precision, as used in~\citetalias{Fouvry.Prunet.2022}.
However, here we exercise the empirical frequency computation from \orbitalelements\ (Appendix~\ref{app:ComputingFrequencies}) for our tests of the isochrone potential.
In what follows in this section, we use our numerical calculations of the frequencies to reproduce earlier results, testing the \ROI\@ and ${\ell\!=\!1}$ damped mode calculations. 
The equations for the isochrone model potential and \DFs\ can be found in~\citetalias{Fouvry.Prunet.2022}.

%%%%%%%%%%%%%%%%%%%%
\subsection[l=2 modes -- Radial Orbit Instability]{${\ell\!=\!2}$ modes -- Radial Orbit Instability} 
\label{app:IsochroneROI}
%%%%%%%%%%%%%%%%%%%%

%%%%%%%%%%%%%%%%%%%%
\input{tables/TableC1}
%%%%%%%%%%%%%%%%%%%%

The \ROI\@ in the isochrone model with ${\rra/\brc\!=\!1.0}$, with $\brc$ the isochrone scale radius, has become a regular test for different aspects of the matrix method implementation \citep{Hamilton.etal.2018,Fouvry.etal.2021,Rozier.etal.2022,Lilley.vandeVen.2023}.
Using the radially-anisotropic \DF\ given in equation~(G12) of \citet{Fouvry.etal.2021}, we compute the growth rate for the fiducial ${\rra/\brc\!=\!1.0}$ case. 
We find a fastest growing mode with ${\gammaM/\Omega_{0}[\rra/\brc\!=\!1.0]\!=\!0.0236\!\pm\!0.0015}$.
This matches the result of~\citetalias{Fouvry.Prunet.2022} at the 2 per cent level, and \citet{Saha.1991} at the 4 per cent level. 
Additionally, the shape of the recovered isochrone \ROI\@ mode matches that of \citetalias{Fouvry.Prunet.2022}, which itself matches the mode shape of \citet{Saha.1991}.

Given the efficiency of \linearresponse\, we are able to test a range of configuration parameters to determine how results may vary with control parameters.
This allows us to establish some measure of uncertainty, as quoted in the above paragraph.
In Table~\ref{tab:ROIResults}, we report the growth rate of the fastest growing mode with various control parameters.

Previous studies have investigated the dependence of the \ROI\ w.r.t.\ the normalised anisotropy radius, $\rra/\brc$.
A naive application of our libraries and the \DF\ given in \citet{Fouvry.etal.2021} suggests that the isochrone stability boundary is ${\rra/\brc\!>\!2}$.
However, in examining the results of our methodology, we discovered a few numerical pitfalls that could alter results.
First, the integration domain should be appropriately limited: as $\rra/\brc$ decreases, regions of the $(J_r,L)$ plane become unpopulated \citep[see, e.g., figure~{4} in ][]{Hamilton.etal.2018}.
Second, the gradient of the \DF\ becomes extremely large at the ${Q\!=\!0}$ boundary, such that numerical estimates readily break down.
To circumvent this, \citet{Hamilton.etal.2018} applied a smoothing to the radially-anisotropic form of the \DF\, hence considering the smoothed \DF\ ${\tilde{f}\!=\!\exp(-\zeta/Q)f}$. 
We find here that different values of $\zeta$, the smoothing length, result in significantly different estimates for the growth rate, at fixed ${\rra/\brc}$.
We do not address either of these shortcomings in the present work.
As a result, obtaining a clear and reliable stability transition radius in $\rra/\brc$ remains out of reach.
We note that even though the growth frequency may be mis-estimated, the shape of the predicted modes is converged.

For completeness, we finally report literature values of the stability boundary in ${\rra/\brc}$.
\citet{May.Binney.1986} estimated an instability threshold at the half-mass radius of ${\rra/\brc\!=\!1.67}$.
\citet{Merritt.1988} then identified an instability threshold of ${\rra/\brc\!=\!1.9}$, from simulations using a harmonic code~\citep{Merritt.1987}.
\citet{Saha.1991} performed new simulations using the same harmonic code from \citet{Merritt.1987}, producing qualitative, but not quantitative, agreement in the trend of growth rates.
The linear response calculations in \citet{Saha.1991}, using a tailored basis, suggested that unstable modes may be found for ${\rra/\brc\!\le\!4}$.
Future work may return to the question of stability with improved numerical treatment.

%%%%%%%%%%%%%%%%%%%%
\subsection[l = 1 modes]{${\ell\!=\!1}$ modes}
\label{app:IsochroneISO}
%%%%%%%%%%%%%%%%%%%%

While searches for instabilities have been fairly commonplace in dynamical modelling, searches for damped modes in stellar clusters have not equally flourished.
Most of this owes to the challenge of calculating the linear response of a system in the lower-half plane.
\citet{Weinberg.1994} identified ${\ell\!=\!1}$ damped modes in \citet{King.1966} models using linear response theory.
\cite{Weinberg.1994} also compared this prediction with perturbed $N$-body simulations, finding qualitative agreement.
\citet{Heggie.Breen.Varri.2020} measured the same ${\ell\!=\!1}$ mode in $N$-body simulations of a King model.

Extending the tools developed for the isochrone cluster, \citetalias{Fouvry.Prunet.2022} measured a damped mode for the isotropic isochrone model at ${\omegaM/\Omega_{0}\!=\!0.0143\!-\!0.00142\ri}$.
As a first test, we reproduced this prediction using all the machinery from \linearresponse\@.
Table~\ref{tab:ISOResults} lists results for a set of validation runs used to search for the ${\ell\!=\!1}$ damped mode.
There, we denote the set of parameters from~\citetalias{Fouvry.Prunet.2022}, for which we recover ${ \omegaM / \Omega_0 \!=\! 0.0143 \!-\! 0.00143\ri }$, in tight agreement with~\citetalias{Fouvry.Prunet.2022}.

However, we noted that varying the control parameters of \linearresponse\ results in a drift of the mode's frequency. 
The middle column of Table~\ref{tab:ISOResults} lists the results for different runs. 
As the number of resonances considered increases (controlled by $\texttt{n1max}$), the predicted complex frequency of the ${\ell\!=\!1}$ mode shifts towards the origin of the complex plane.
On the bright side though, when holding other parameters fixed and varying only the scale length of the basis (\texttt{rb}), we find that the result is essentially converged: the dilatation of the basis can be absorbed in this problem.

%%%%%%%%%%%%%%%%%%%%
\input{tables/TableC2}
%%%%%%%%%%%%%%%%%%%%

For the calculations considered here, the cluster's pure neutral translation mode \citep[the ${\ell\!=\!1}$ mode at ${\omegaM\!=\!0\!+\!0\ri}$, see][]{Weinberg.1989,Murali.1999} is not converged, and appears as a `false' damped mode.
This seems to be the mode we are recovering here. 
We do not find evidence for any other ${\ell\!=\!1}$ modes in the isotropic isochrone model.

False modes are more likely to manifest themselves in the lower half-plane due to the Landau prescription employed in computing the integral presented in equation~(\ref{def_D}).
Indeed, the divergence of $P_{k} (\omega)$ (equation~\ref{def_D}) only contributes in the lower-half of the complex frequency plane.

Regrettably, due to numerical noise in the calculation, we cannot explore sufficiently low regions in the complex plane to identify authentic damped modes that might exist below our imposed limits.
Hopefully, upcoming numerical enhancements should enable us to delve deeper into the complex frequency plane.
Finally, let us stress that the presence of very damped modes  are unlikely to be relevant in Nature. 
Indeed, in a ${N\!=\!10^{5}}$ globular cluster, the evolution of fluctuations can be taken to be collisionless only for large enough amplitude, hence making globular clusters unable to ``resolve'' very damped linear modes.

%%%%%%%%%%%%%%%%%%%%
\section{The Plummer Sphere}
\label{app:Plummer}
%%%%%%%%%%%%%%%%%%%%

Unlike the isochrone sphere, the Plummer sphere does not have analytical expressions for its orbital frequencies. 
We therefore use the tools in \orbitalelements\ to compute the frequencies from the potential (and its two derivatives), as well as the resonantly-aligned ${(u,v)}$ coordinates.
Here, the versatility of \linearresponse\ shines: the user only needs to define the \DF\@.

The \citet{Plummer.1911} potential is given by 
\begin{equation}
    \psi(r) = - GM / \sqrt{\brc^2+r^2} ,
    \label{eq:Plummer_Pot}
\end{equation}
with $\brc$ the scale radius. 
The frequency scale for the Plummer sphere, $\Omega_0$, is given by
\begin{equation}
    \Omega_0 = \Omega_1(r\to0) = 2\sqrt{ GM / \brc}.
    \label{eq:plummeromega0}
\end{equation}
The isotropic \DF\ is
\begin{equation}
    F(E) = \frac{24\sqrt{2}}{7\pi^3}\left(\frac{E}{E_0}\right)^{7 / 2},
\end{equation}
with the energy scale $E_0=-GM/\brc$.

To introduce radial anisotropy, we use the \citet{Osipkov.1979,Merritt.1985} method to find a self-consistent \DF\@. 
The dimensionless quantity
\begin{equation}
    Q = \frac{1}{E_0}\left(E+\frac{L^2}{2 \rra^2}\right)
\end{equation}
is the single parameter in the \DF\@, which reads\footnote{Equation~{(5)} in~\cite{Breen.Varri.Heggie.2017} does not include a square on the middle term in brackets. This appears to be a typesetting error, as the \texttt{PlummerPlus.py} software (see Appendix~\ref{app:NBody}) includes this square.}
\begin{equation}
    F(Q) = \frac{24\sqrt{2}}{7\pi^3} \, Q^{7 / 2} \,\left[1-\frac{1}{\rra^2}+\frac{7}{16\rra^2Q^2}\right].
\end{equation}

%%%%%%%%%%%%%%%%%%%%
\section[N-Body Technique]{$N$-Body Technique}
\label{app:NBody}
%%%%%%%%%%%%%%%%%%%%

To compare our linear response theory predictions with $N$-body simulations, we generate realisations of the Plummer cluster and evolve them for several hundred dynamical times. 
The results are discussed in Section~\ref{sec:SimulationResults}. 
This appendix gives a brief description of the $N$-body method, and the parameters of the $N$-body runs (see Table~\ref{tab:Nbodyparams}).\footnote{Any unspecified parameters are taken as software defaults.}

%%%%%%%%%%%%%%%%%%%%
\input{tables/TableE1}
%%%%%%%%%%%%%%%%%%%%

The initial conditions are generated using \plummerplus\ \citep{Breen.Varri.Heggie.2017}. 
We use a lightly modified version of \texttt{PlummerPlus.py}, specifically to produce outputs that interface directly with \EXP\@ \citep{Petersen.Weinberg.Katz.2022}, a basis function expansion $N$-body code.\footnote{The code we use is available at \href{https://github.com/michael-petersen/PlummerPlus}{https://github.com/michael-petersen/PlummerPlus}, forked from \href{https://github.com/pgbreen/PlummerPlus}{the original}.}
The initial conditions are centred by the median positions and velocities in each dimension after sampling.

Simulations are run using \EXP\@. 
For the basis, we use the Sturm--Liouville solution from \EXP\@.
In the case of the Plummer model, the Sturm--Liouville solver in \EXP\ results in the same basis as the analytical \citet{CluttonBrock.1973} basis (cf.\ Appendix~\ref{app:ComputingBasis}). 

Conveniently, \EXP\ uses virial units, making direct comparison between the simulation and linear response calculations straightforward. 
We run simulations for 400 time units, with a maximum possible timestep of 0.2. 
In practice, \EXP\ employs a block-step technique for subdividing timesteps in powers of 2. We allow a maximum of 7 divisions, corresponding to a minimum timestep of ${\sim\!1.5\!\times\!10^{-3}}$. 
The median timestep for particles is ${0.2/2^3\!=\!0.025}$.

As is the default behaviour in \EXP\@, the centre of the expansion is pinned at the inertial centre, leaving to the basis to resolve any excursions. 
As the maximum excursion of the peak density is well within a scale radius of the simulated clusters, we do not expect any bias \citep[cf.][]{Petersen.Weinberg.Katz.2022}.

For computational expediency, we draw a relatively small number of particles per cluster,\footnote{No tests w.r.t.\ $N$ were performed, though we expect that the measurement uncertainty will decrease like $\sqrt{N}$.}
but realise six clusters per \DF\ to estimate the average response.
For the ROI measurement in Fig.~\ref{fig:Figure3}, we shift time such that the peak potential fluctuations line up.

To efficiently estimate the magnitude of the potential fluctuations, ${ \Delta \psi }$, we first compute the self-gravity in the non-axisymmetric harmonics. 
Given that the square of the \EXP\ coefficients is the self-gravity \citep[cf.\ section~{1} of][]{Petersen.Weinberg.Katz.2022}, and the magnitude of the amplitudes encodes $\Delta \psi$, we may efficiently measure the potential fluctuations in either self-gravity or $\Delta\psi$, simply using the output products of the simulation
\begin{equation}
    \Delta \psi^{\ell} = \sum_p a_p^{\ell} \, \psi_{p}^{\ell},
\end{equation}
where $a_p^{\ell}$ is the coefficient of the given basis function $\psi_p^{\ell}$. 
This is used in Figs.~\ref{fig:Figure3} and \ref{fig:Figure4}.

To compute the energy in a given harmonic, we compute the sum square of the amplitudes of all radial orders in a given harmonic,
\begin{equation}
    E_{\ell} = \sum_{p} \big( a_p^{\ell} \big)^{2} .
    \label{eq:def_Al}
\end{equation}
To provide a natural scaling, harmonics with ${\ell\!>\!0}$ are normalised by the monopole to define
\begin{equation}
\tE_{\ell} = E_{\ell} / E_{\ell = 0} .
\label{eq:def_tE}
\end{equation}
In Fig.~\ref{fig:Figure3}, we measure the slope of ${\ln\!\left(\tilde{E}_2\right)}$ to obtain the growth rate for the ROI simulations.
The oscillation frequency of a mode, determined from a given coefficient for a harmonic $\ell$, is computed as
\begin{equation}
    \OmegaM = \arctan\left( \ImPart[a_p^{\ell}] / \RePart[a_p^{\ell}] \right) .
\end{equation}
Measurements based on the coefficients are not perfect, in that both truncation noise and particle noise may bias the coefficients. 
Fortunately, these errors are generally small. 
In the context of the present simulations, reasonable estimates for the error on ${\Delta\psi^{\ell}/\psi_0}$ (where $\psi_0$ is the equilibrium model potential) and $\tE_{\ell}$ are of order 1\%. 

%%%%%%%%%%%%%%%%%%%%
\section{Razor-thin discs}
\label{app:Discs}
%%%%%%%%%%%%%%%%%%%%

%%%%%%%%%%%%%%%%%%%%
\subsection{{2D} bi-orthogonal bases}
\label{app:2Dbasis}
%%%%%%%%%%%%%%%%%%%%

Due to axial symmetry, it is natural to contemplate basis elements that are harmonically decoupled.
Potential-density pairs for a disc then take the form ${[ U_{p}^{\ell} (r) \, \re^{\ri\ell\phi},\, D_{p}^{\ell} (r) \, \re^{\ri\ell\phi} ]}$ where ${(r,\phi)}$ are the usual polar coordinates and $p$ (resp.~$\ell$) stands for the radial (resp.~azimuthal) number.
As in the spherical case, a basis is then fully prescribed by its radial potential functions, $U_{p}^{\ell}$.
In \astrobasis\@, we implemented the $2D$ basis from \cite{CluttonBrock.1972} (infinite radial support) and \cite{Kalnajs.1976} (finite radial support), with a careful treatment of the normalisation.
Naturally, it would be of  interest to implement tailored basis functions \citep{Petersen.Weinberg.Katz.2022,Lilley.vandeVen.2023}.
This is left for future work.

%%%%%%%%%%%%%%%%%%%%
\input{tables/TableF1}
%%%%%%%%%%%%%%%%%%%%

%%%%%%%%%%%%%%%%%%%%
\input{tables/TableF2}
%%%%%%%%%%%%%%%%%%%%

%%%%%%%%%%%%%%%%%%%%
\subsection{Zang discs}
\label{app:DiscsDF}
%%%%%%%%%%%%%%%%%%%%

The stability analysis of tapered Mestel discs was pioneered by \cite{Zang.1976} and subsequently generalised by \cite{Evans.Read.1998.II}.
We refer to section~{4.1} of \cite{Fouvry.etal.2015} for a concise presentation of these discs, their \DFs\ and potentials.

We place ourselves in the same unit system as in \cite{Fouvry.etal.2015}, hence setting ${V_0\!=\!G\!=\!R_{\ri}\!=\!1}$. 
Within this unit system, a natural scale frequency is ${ \Omega_0\!=\!V_0/R_{\ri}\!=\!1 }$. Here, $\Omega_0$ does not stand for the (infinite) central radial frequency.
As in \cite{Zang.1976}, we consider ``fully'' self-gravitating discs (active fraction ${\xi\!=\!1}$) with radial velocity dispersion such that ${q\!=\!6}$, while varying the inner taper exponent $\nu_{\rt}(=\!N)$.
The outer taper is taken sufficiently far (${R_{\mathrm{o}}\!=\!11.5}$) and sharp (${\mu_{\rt}\!=\!M\!=\!5}$) not to affect the populated regions, hence not affecting the disc's response.

For numerical purpose in \orbitalelements\@, we consider a softened Mestel potential
\begin{equation}
    \label{eq:truncated_Mestel_potential}
    \psiM(r) = \half V_0^2 \ln [\varepsilon^2 + (r/R_0)^2],
\end{equation}
to prevent singularities for exactly radial orbits.
In practice, $\varepsilon$ is taken sufficiently small (${\varepsilon\!=\!10^{-5}}$; ${R_0\!=\!1}$) compared to the \DF\@'s inner taper radius ${R_{\ri}\!=\!1}$.
We checked that this softening had no impact on the predicted linear response.
Table~\ref{tab:LRdiscsparams} summarises the parameters used in \linearresponse\@ for the disc ``Fiducial'' numerical applications.
In Table~\ref{tab:ZangVariance}, we report the location of the fastest growing mode for the ${N\!=\!4}$ Zang disc with various control parameters. The ``Fiducial'' run is well converged.

%% file: tables/TableA1.tex
\begin{table*}
\centering
  \begin{tabular}{l|l|l|l|}
Parameter & Datatype & Default & Description  \\
    \hline
    \hline
    \texttt{EDGE}             & \float & 0.01 & Tolerance in $w$ before switching to pericentre and apocentre expansions in equation~\eqref{eq:Theta} \\ % in units of w
    \texttt{NINT}             & \integer & 32 & Number of steps for the Simpson's ${1/3}$ integration rule in $w$ to compute equations~\eqref{eq:AlphaBeta_w}\\
    \texttt{TOLECC}           & \float &0.01 & Tolerance in eccentricity before expanding the energy, actions and frequencies from circular or radial orbits (Section~\ref{app:ForwardMappings})\\ % in units of eccentricity
    \texttt{TOLA}             & \float &0.01 & Tolerance in semi-major axis before expanding the energy, actions and frequencies from $r=0$ (Section~\ref{app:ForwardMappings})\\
    \texttt{ITERMAX}          & \integer & 100 & Number of Newton--Raphson descent steps for backward mappings (Section~\ref{app:BackwardMappings})\\
    \texttt{inv$\varepsilon$} & \float &1e-12 & Target accuracy on Newton--Raphson inversions (Section~\ref{app:BackwardMappings})\\
    \texttt{$\Omega_0$}       & \float &1.0 & Frequency normalisation scale \\
    \texttt{da}               & \float &1.e-4 & Step size for numerical derivative w.r.t.\ the semi-major axis (Section~\ref{app:BackwardMappings})\\
    \texttt{de}               & \float &1.e-4 & Step size for numerical derivative w.r.t.\ the eccentricity (Section~\ref{app:BackwardMappings})\\
    \texttt{rc}             & \float & 1.0 & Characteristic scale radius \\
    \texttt{rmin}             & \float & 0.0 & Inner truncation radius for the resonance variables $(u,v)$ (Section~\ref{app:ResonancesVariables}) \\
    \texttt{rmax}             & \float & \texttt{Inf} & Outer truncation radius for the resonance variables  $(u,v)$ (Section~\ref{app:ResonancesVariables}) \\
\end{tabular}
  \caption{Summary of the control parameters in \orbitalelements\@. Further description is given throughout Appendix~\ref{app:ComputingFrequencies}.}
  \label{tab:OEparams}
\end{table*}

%% file: tables/TableA2.tex
\begin{table}
\setlength{\tabcolsep}{3pt}
\centering
  \begin{tabular}{l|l|l|l|}
Parameter & Datatype & Default & Description  \\
    \hline
    \hline
    \texttt{rb}             & \float & 1.0 & Radial scale of the basis functions \\
    \texttt{G}             & \float & 1.0 & Gravitational constant in Poisson equation \\
    \texttt{nradial}             & \integer & none & Maximum radial order of the expansion \\
\end{tabular}
  \caption{Summary of the control parameters in \astrobasis\@. Further description is given in the text.}
  \label{tab:ABparams}
\end{table}

%% file: tables/TableA3.tex
\begin{table*}
\centering
  \begin{tabular}{l|l|l|l|}
Parameter & Datatype & Default & Description  \\
    \hline
    \hline
    \texttt{Ku}           & \integer & 200            & Number of points to sample per resonance in $u$ coordinate \\
    \texttt{Kv}           & \integer & 200            & Number of points to sample per resonance in $v$ coordinate \\
    \texttt{Kw}           & \integer & 200            & Number of points to sample per orbit to compute Fourier transform of basis elements (equation~\ref{eq:Wln})\\
    \texttt{ADAPTIVEKW}   & \bool    & \texttt{false} & If \texttt{true}, automatically scale number of \texttt{Kw} points to optimise calculation (equation~\ref{eq:ADAPTIVEKW}) \\
    \texttt{VMAPN}        & \integer & 1              & Exponent in the ${v\!\mapsto\!v'}$ mapping from equation~\eqref{eq:VMAPN}. Larger values will sample more finely near $\vnm$ \\
    \texttt{lharmonic}    & \integer & 2              & Harmonic $\ell$ to be considered \\
    \texttt{n1max}        & \integer & 10              & Maximum radial resonance number $n_1$ to consider in Eq.~\eqref{eq:M_int_u} (azimuthal number $n_2$ constrained by \texttt{lharmonic}) \\
\end{tabular}
  \caption{Summary of the control parameters in \linearresponse\@. Further description is given throughout Appendix~\ref{app:ComputingLR}.}
  \label{tab:LRparams}
\end{table*}

%% file: tables/TableC1.tex
\begin{table}
\centering
  \begin{tabular}{l|l|l|}
    Run  & Isochrone $\gammaM / \Omega_{0}$ & Plummer $\gammaM / \Omega_{0}$ \\
    \hline
    \hline
Fiducial                  & 0.0236 & 0.0477 \\
\texttt{Ku}$\times$2      & 0.0247 & 0.0477 \\
%\texttt{Ku}$\times$4      & fails! &  \\
\texttt{Kv}$\times$2      & 0.0243 & 0.0477 \\
\texttt{Kw}$\times$2      & 0.0238 & 0.0477 \\
\texttt{nradial}/2        & 0.0238 & 0.0470 \\
\texttt{n1max}$=$1        & 0.0238 & 0.0470 \\
\texttt{n1max}$\times$2   & 0.0237 & 0.0478 \\
\texttt{n1max}$\times$4   & 0.0238 & 0.0479 \\
\texttt{rb}$/$5           & 0.0236 & 0.0476 \\
\texttt{rb}$\times$4 \citepalias{Fouvry.Prunet.2022} & 0.0238 & 0.0477 \\
\texttt{VMAPN}=2          & 0.0251 & 0.0477 \\
\end{tabular}
  \caption{Predictions for the \ROI\@ ${\ell\!=\!2}$ growth rates in the isochrone (${\rra/\brc\!=\!1.0}$) and Plummer (${\rra/\brc\!=\!0.75}$) clusters. The configurations with increasing \texttt{n1max} are the most computationally expensive: \texttt{n1max}$\times2$ requires 122 resonances, while \texttt{n1max}$\times4$ requires 242 resonances.
  }
  \label{tab:ROIResults}
\end{table}

%% file: tables/TableC2.tex
\begin{table}
\centering
  \begin{tabular}{l|l|l|}
    Run  & Isochrone $\omegaM / \Omega_{0}$ & Plummer $\omegaM / \Omega_{0}$ \\
    \hline
    \hline
Fiducial                & $0.0143 - 0.0014\ri$ & $0.0247 - 0.0004\ri$ \\
\texttt{Ku}$\times$2    & $0.0150 - 0.0022\ri$ & $0.0247 - 0.0004\ri$ \\
%\texttt{Ku}$\times$4    & $0.0142 - 0.0012\ri$ & $0.0247 - 0.0004\ri$ \\
\texttt{Kv}$\times$2    & $0.0143 - 0.0014\ri$ & $0.0247 - 0.0004\ri$ \\
\texttt{Kw}$\times$2    & $0.0143 - 0.0014\ri$ & $0.0247 - 0.0004\ri$ \\
\texttt{nradial}/2      & $0.0143 - 0.0014\ri$ & $0.0246 - 0.0004\ri$ \\
\texttt{n1max}$\times$2 & $0.0084 - 0.0005\ri$ & $0.0124 - 0.0001\ri$ \\
\texttt{n1max}$\times$4 & $0.0048 - 0.0002\ri$ & $0.0060 - 0.0001\ri$ \\
\texttt{rb}$/$5         & $0.0143 - 0.0014\ri$ & $0.0247 - 0.0004\ri$ \\
\texttt{rb}$\times$4 \citepalias{Fouvry.Prunet.2022} &$0.0143 - 0.0014\ri $ & $0.0247 - 0.0004\ri$\\
\texttt{VMAPN}$=2$      & $0.0143 - 0.0014\ri$ & $0.0246 - 0.0004\ri$ \\
\end{tabular}
  \caption{Predictions for ${\ell\!=\!1}$ mode locations in the complex plane. The search space is ${ \RePart (\omega) / \Omega_{0} \!\in\! [0.0,0.1] }$ and ${ \ImPart [\omega] / \Omega_{0} \!\in\! [-0.01,0.05] }$.
  In general, the more resonances are considered,
  the more the complex frequency of the recovered (damped) ${ \ell \!=\! 1 }$ mode
  converges to the frequency of the (neutral) translation mode
  ${ \omegaM \!=\! 0 \!+\! 0 \ri }$.
  }
  \label{tab:ISOResults}
\end{table}

%% file: tables/TableE1.tex
\begin{table}
\setlength{\tabcolsep}{3pt}
\centering
  \begin{tabular}{l|l|l|}
Parameter & Value & Description  \\
    \hline
    \hline
    \texttt{lmax}     & 6 & Number of azimuthal harmonics \\
    \texttt{nmax}     & 24 & Number of radial basis functions \\
    \texttt{N}        & 262\,144 & Number of equal-mass particles \\
    \texttt{multistep} & 7 & Maximum subdivision of the timestep \\
    \texttt{dt}       & 0.2 [virial] & With \texttt{multistep}, minimum timestep is $0.2/2^7$ \\
    \texttt{nsteps} & 2\,000 & With \texttt{dt}, total run time is 400 virial units
\end{tabular}
  \caption{Summary of the control parameters for then \EXP\ $N$-body runs.}
  \label{tab:Nbodyparams}
\end{table}

%% file: tables/TableF1.tex
\begin{table}
\setlength{\tabcolsep}{3pt}
\centering
  \begin{tabular}{l|l|l|}
     Library & Parameter & Value  \\
    \hline
    \hline
    \orbitalelements\@ & \texttt{rmin} & 0.2 \\
    \astrobasis\@      & Basis & \cite{CluttonBrock.1972} \\
                     & \texttt{rb} & 5.0 \\
                     & \texttt{nradial} & 100 \\
    \linearresponse\@  & \texttt{VMAPN} & 2 \\
\end{tabular}
  \caption{Summary of the control parameters for the disc's linear response computations.
  Unspecified parameters are taken as defaults (Tables \ref{tab:OEparams}--\ref{tab:LRparams}).}
  \label{tab:LRdiscsparams}
\end{table}

%% file: tables/TableF2.tex
\begin{table}
\centering
  \begin{tabular}{l|l|l|}
    Run  & $\OmegaM / \Omega_0$ & $\gammaM / \Omega_0$ \\
    \hline
    \hline
Fiducial                  & 0.878 & 0.126 \\
\texttt{Ku}$\times$2      & 0.878 & 0.126 \\
\texttt{Kv}$\times$2      & 0.878 & 0.126 \\
\texttt{Kw}$\times$2      & 0.878 & 0.126 \\
\texttt{nradial}/2        & 0.878 & 0.126 \\
\texttt{n1max}$=$1        & 0.842 & 0.028 \\
\texttt{n1max}$\times$2   & 0.879 & 0.127 \\
\texttt{n1max}$\times$4   & 0.879 & 0.127 \\
\texttt{rb}$/$5           & 0.878 & 0.126 \\
\texttt{rb}$\times$4      & 0.878 & 0.128 \\
\texttt{VMAPN}=1          & 0.885 & 0.135 \\
\texttt{VMAPN}=1, \texttt{Kv}$\times$2 & 0.881 & 0.128 \\
\texttt{VMAPN}=1, \texttt{Kv}$\times$4 & 0.879 & 0.127 \\
\end{tabular}
  \caption{Predictions for the ${N\!=\!4}$ Zang disc most unstable mode: oscillation frequencies and growth rates when varying the control parameters. When considering enough resonances (\texttt{n1max} sufficiently large), the predictions are strongly consistent. Spreading the point more evenly along the resonance line (via setting ${\texttt{VMAPN}\!=\!2}$ in equation~\ref{eq:VMAPN}) prove particularly useful in reducing the number of integration points \texttt{Kv} needed to reach convergence, hence saving computational time and reducing memory allocation. 
  }
  \label{tab:ZangVariance}
\end{table}